\documentclass[multphys,vecphys]{svmult}

\usepackage{amsmath}
\usepackage{amssymb,amsfonts}
\usepackage{bm}
\usepackage[mathcal]{euscript}
\usepackage{graphicx}
\usepackage{psfrag}
\usepackage{subfigure}

\usepackage{makeidx}         
\usepackage{graphicx}        
\usepackage{multicol}        
\usepackage[bottom]{footmisc}

\usepackage{cite}

\newcommand{\comment}[1]{}

\newcommand{\ie}{\textit{i.e.}}

\newcommand{\goodgap}{%
	\hspace{8pt}}

\newcommand{\mathnotation}[2]{\newcommand{#1}{\ensuremath{#2}}}

\newcommand{\Order}[1]{\ensuremath{\mathcal{O}\!\l(#1\r)}}

\newcommand{\Expect}[1]{\overline{#1}}		

\DeclareMathOperator{\Tr}{Tr}

\renewcommand{\l}{\left}			
\renewcommand{\r}{\right}			
\mathnotation{\pd}{\partial}			
\mathnotation{\ee}{{\mathrm e}}			
\mathnotation{\imi}{\mathrm{i}}			
\mathnotation{\ldef}{\mathrel{\raisebox{.069ex}{:}\!\!=}}
\mathnotation{\rdef}{\mathrel{=\!\!\raisebox{.069ex}{:}}}
\mathnotation{\dint}{\,{\mathrm{d}}}		

\mathnotation{\grad}{\nabla}			
\renewcommand{\div}{\grad\cdot}			
\mathnotation{\curl}{\grad\times}		
\mathnotation{\lapl}{\nabla^2}			

\mathnotation{\Gauss}{G}			
\mathnotation{\Prob}{P}				
\mathnotation{\Cramer}{\mathcal{S}}		
\mathnotation{\spt}{{\mathrm{sp}}}		
\mathnotation{\sptt}{{\mathrm{0}}}		

\renewcommand{\time}{t}				
\mathnotation{\stime}{s}			
\mathnotation{\iter}{n}				
\mathnotation{\iterj}{m}			
\mathnotation{\xc}{x}				
\mathnotation{\xv}{{\bm{\xc}}}			
\mathnotation{\velc}{v}
\mathnotation{\velv}{{\bm{\velc}}}

\mathnotation{\Var}{\mathrm{Var}}		
\mathnotation{\kc}{k}				
\mathnotation{\kv}{{\bm{\kc}}}			
\mathnotation{\qv}{{\bm{q}}}			
\mathnotation{\kzc}{k_0}			
\mathnotation{\kzv}{{\bm{k}}_0}			
\mathnotation{\kzct}{{\widetilde k}_0{}}	
\mathnotation{\Tf}{\mathcal{T}}			
\mathnotation{\Id}{\mathrm{Id}}			

\mathnotation{\Diff}{\kappa}			
\mathnotation{\visc}{\nu}			
\mathnotation{\fwid}{\ell}			
\mathnotation{\shrlen}{\chi}			

\mathnotation{\fluc}{\beta}			

\mathnotation{\sdim}{d}				

\mathnotation{\Pe}{\mathrm{Pe}}			
\mathnotation{\Sc}{\mathrm{Sc}}			

\newcommand{\adeq}{advection--diffusion equation}
\newcommand\etal{\mbox{\textit{et al.}}}

\begin{document}

\title*{Scalar Decay in Chaotic Mixing}
\author{Jean-Luc Thiffeault}
\institute{Department of Mathematics, Imperial College
  London, 
  United Kingdom
\texttt{jeanluc@imperial.ac.uk}}
%
%
\maketitle

\begin{abstract}
I review the \emph{local theory} of mixing, which focuses on infinitesimal
blobs of scalar being advected and stretched by a random velocity field.  An
advantage of this theory is that it provides elegant analytical results.  A
disadvantage is that it is highly idealised.  Nevertheless, it provides
insight into the mechanism of chaotic mixing and the effect of random
fluctuations on the rate of decay of the concentration field of a passive
scalar.
\end{abstract}


\section{Introduction}
\label{sec:intro}

The equation that is in the spotlight is the \emph{\adeq}
\begin{equation}
  \pd_\time\theta + \velv\cdot\grad\theta = \Diff\lapl\theta
  \label{eq:adeq}
\end{equation}
for the time-evolution of a distribution of concentration~$\theta(\xv,\time)$,
being \emph{advected} by a velocity field~$\velv(\xv,\time)$, and
\emph{diffused} with diffusivity~$\Diff$.  The concentration~$\theta$ is
called a \emph{scalar} (as opposed to a vector).  We will restrict our
attention to incompressible velocity fields, for which~$\div\velv=0$.  For our
purposes, we shall leave the exact nature of~$\theta$ nebulous: it could be a
temperature, the concentration of salt, dye, chemicals, isotopes, or even
plankton.  The only assumption for now is that this scalar is \emph{passive},
which means that its value does not affect the velocity field~$\velv$.
Clearly, this is not strictly true of some scalars like temperature, because a
varying buoyancy influences the flow, but is often a good approximation
nonetheless.

The \adeq\ is linear, but contrary to popular belief that does not mean it is
simple!  Because the velocity (which is regarded here as a given vector field)
is a function of space and time, the advection term (the second term
in~\eqref{eq:adeq}) can cause complicated behaviour in~$\theta$.  Broadly
speaking, the advection term tends to create sharp gradients of~$\theta$,
whilst the diffusion term (the term on the right-hand side of~\eqref{eq:adeq})
tends to wipe out gradients.  The evolution of the concentration field is thus
given by a delicate balance of advection and diffusion.

The advection term in~\eqref{eq:adeq} is also known as the \emph{stirring}
term, and the interplay of advection and diffusion is often called
\emph{stirring and mixing}.  As we shall see, the two terms have very
different r\^oles, but both are needed to achieve an efficient mixing.

To elicit some broad features of mixing, we will start by deriving some
properties of the \adeq.  First, it conserves the total quantity of~$\theta$.
If we use angle brackets to denote the average of~$\theta$ over the fixed
domain of interest~$V$, \ie
\begin{equation*}
  \l\langle \theta\r\rangle \ldef \frac{1}{V}\int_V\theta\dint V,
\end{equation*}
then we find diretly from~\eqref{eq:adeq} that
\begin{equation}
  \pd_\time\l\langle\theta\r\rangle + \l\langle\velv\cdot\grad\theta\r\rangle
  = \Diff\l\langle\lapl\theta\r\rangle.
  \label{eq:adeqavg}
\end{equation}
Because the velocity field is incompressible, we have
\begin{equation*}
  \velv\cdot\grad\theta = \div(\theta\,\velv),
\end{equation*}
and also~\hbox{$\lapl\theta = \div(\grad\theta)$}.  Thus, we can use the
divergence theorem to write~\eqref{eq:adeqavg} as
\begin{equation}
  \pd_\time\l\langle\theta\r\rangle =
  -\frac{1}{V}\int_S\theta\,\velv\cdot\hat{\bm{n}}\dint S
  + \Diff\,\frac{1}{V}\int_S\grad\theta\cdot\hat{\bm{n}}\dint S,
  \label{eq:adeqavg2}
\end{equation}
where~$S$ is the surface bounding~$V$, and~$\dint S$ is the element of area,
and~$\hat{\bm{n}}$ outward-pointing normal to the surface.  For a closed flow,
two possibilities are now open to us: (i) the domain~$V$ is periodic; or (ii)
$\velv$ and $\grad\theta$ are both tangent to the surface~$S$.  In the first
case, the terms on the right-hand side of~\eqref{eq:adeqavg2} vanish because
boundary terms always vanish with periodic boundary conditions (a bit
tautological, but true!).  In the second case, both~$\velv\cdot\hat{\bm{n}}$
and~$\grad\theta\cdot\hat{\bm{n}}$ vanish.  Either way,
\begin{equation}
  \pd_\time\l\langle\theta\r\rangle = 0
  \label{eq:adeqavgcons}
\end{equation}
so that the mean value of~$\theta$ is constant.  Since~$V$ is constant, this
also implies that the total amount of~$\theta$ is conserved.  The second set
of boundary conditions we used implies that there is no fluid flow or flux
of~$\theta$ through the boundary of the volume.  It is thus natural that the
total~$\theta$ is conserved!  For periodic boundary conditions, whatever
leaves the volume re-enters on the other side, so it also makes sense
that~$\theta$ is conserved.  Because of~\eqref{eq:adeqavgcons}, and because we
can always add a constant to~$\theta$ without changing its evolution (only
derivatives of~$\theta$ appear in~\eqref{eq:adeq}), we will always choose
\begin{equation}
  \l\langle\theta\r\rangle = 0
  \label{eq:zeroavg}
\end{equation}
without loss of generality.  In words: the mean of our scalar vanishes
initially, so by~\eqref{eq:adeqavgcons} it must vanish for all times.

Now let's look at another average of~$\theta$: rather than averaging~$\theta$
itself, which has yielded an important but boring result, we average its
square.  The \emph{variance} is defined by
\begin{equation}
  \Var \ldef \langle\theta^2\rangle - \langle\theta\rangle^2,
  \label{eq:Var}
\end{equation}
where the second term on the right vanishes by~\eqref{eq:zeroavg}.  To obtain
an equation for the time-evolution of the variance, we
multiply~\eqref{eq:adeq} by~$\theta$ and integrate,
\begin{equation*}
  \l\langle\theta\,\pd_\time\theta\r\rangle
  + \l\langle\theta\,\velv\cdot\grad\theta\r\rangle
  = \Diff\l\langle\theta\,\lapl\theta\r\rangle.
\end{equation*}
We rearrange on the left and integrate by parts on the right, to find
\begin{equation*}
  \l\langle\l(\pd_\time + \velv\cdot\grad\r)\tfrac{1}{2}\,\theta^2\r\rangle
  = \Diff\l\langle\div(\theta\,\grad\theta)
  - \lvert\grad\theta\rvert^2\r\rangle.
\end{equation*}
Now there are some boundary terms that vanish under the same assumptions as
before, and we get
\begin{equation}
  \pd_\time\Var = -2\Diff\l\langle\lvert\grad\theta\rvert^2\r\rangle.
  \label{eq:Vareq}
\end{equation}
Notice that, once again, the velocity field has dropped out of this averaged
equation.  However, now the effect of diffusion remains.  Moreover, it is
clear that the term on the right-hand side of~\eqref{eq:Vareq} is
\emph{negative-definite} (or zero): this means that the variance always
decreases (or is constant).  The only way it can stop decreasing is
if~$\grad\theta$ vanishes everywhere, that is, $\theta$ is constant in space.
But because we have assumed~$\l\langle\theta\r\rangle=0$, this means
that~$\theta=0$ everywhere.  In that case, we have no choice but to declare
the system to be \emph{perfectly mixed}: there are no variations in~$\theta$
at all anymore.  Equation~\eqref{eq:Vareq} tells us that variance tends to
zero, which means that the system inexorably tends to the perfectly mixed
state, without necessarily ever reaching it.
Variance is thus a useful measure of mixing: the smaller the variance, the
better the mixing.

There is a problem with all this: equation~\eqref{eq:Vareq} no longer involves
the velocity field.  But if variance is to give us a measure of mixing,
shouldn't its time-evolution involve the velocity field?  Is this telling us
that stirring has no effect on mixing?  Of course not, as any coffee-drinker
will testify, whether she likes it with milk or sugar: stirring has a huge
impact on mixing!  So what's the catch?

\label{page:thecatch}
The catch is that~\eqref{eq:Vareq} is not a closed equation for the variance:
the right-hand side involves~$\lvert\grad\theta\rvert^2$, which is not the
same as~$\theta^2$.  The extra gradient makes all the difference.  As we will
see, under the right circumstances the stirring velocity field creates very
large gradients in the concentration field, which makes variance decrease much
faster than it would if diffusivity were acting alone.  In fact, when~$\Diff$
is very small, in the best stirring flows the gradients of~$\theta$ scale
as~$\Diff^{-1/2}$, so that the right-hand side of~\eqref{eq:Vareq} becomes
independent of the diffusivity.  This, in a nutshell, is the essence of
\emph{enhanced mixing}.

Several important questions can now be raised:
\begin{itemize}
\item How fast is the approach to the perfectly-mixed state?
\item How does this depend on~$\Diff$?
\item What does the concentration field look like for long times?
What is its spectrum?
\item How does the probability distribution of~$\theta$ evolve?
\item Which stirring fields give efficient mixing?
\end{itemize}
The answers to these questions are quite complicated, and not fully known.  In
the following sections we will attempt to give some hints of the answers and
give some references to the literature.

This is not meant to be a comprehensive review article, so entire swaths of
the literature are missing.  We focus mainly on \emph{local} or
\emph{Lagrangian theories}, which involve deterministic and stochastic
approaches for quantifying stretching using a local idealisation of the flow.
The essential feature here is that the \adeq\ is solved along fluid
trajectories.  These theories trace their origins to
Batchelor~\cite{Batchelor1959}, who treated constant matrices with slow time
dependence, and Kraichnan, who introduced fast (delta-correlated) time
dependence~\cite{Kraichnan1968,Kraichnan1974}.
Zeldovich~\etal~\cite{Zeldovich1984} approached the problem from the
random-matrix theory angle in the magnetic dynamo context.  More recently,
techniques from large-deviation
theory~\cite{Ott1989,Antonsen1991,Antonsen1995,Antonsen1996} and path
integration~\cite{Shraiman1994,Chertkov1995,Chertkov1997,Shraiman2000,%
Falkovich2001} have allowed an essentially complete solution of the problem.
It is this work that will be reviewed here, as it applies to the decay of the
passive scalar (and not the PDF of concentration or its power spectrum).  We
will favour expediency over mathematical rigour, and try to give a flavour of
what these local theories are about without describing them in detail.

The story will proceed from here as follows: in Section~\ref{sec:linear}
advection of a blob by a linear velocity field is considered, with diffusion
included.  This problem has an exact solution, but it can be made simpler in
the limit of small diffusivity.  Solutions are examined for a straining flow
in two and three dimensions (Sections~\ref{sec:strain} and~\ref{sec:3D}), as
well as a shear flow in two dimensions (Section~\ref{sec:shear}).  Randomness
is added in Section~\ref{sec:rsm}: the strain associated with the velocity
field is assumed to vary, and the consequences of this for a single blob
(Section~\ref{sec:oneblob}) and a large number of blobs
(Section~\ref{sec:manyblobs}) are explored.  Practical implementation is
discussed in Section~\ref{sec:practical}, and a simple model for a micromixer
is analysed in Section~\ref{sec:decayexample}.  Finally, the limitations of
the theory presented herein (Section~\ref{sec:limitations}).

\section{Advection and Diffusion in a Linear Velocity Field}
\label{sec:linear}

We will start by considering what happens to a passive scalar advected by a
linear velocity field.  The overriding advantage of this configuration is that
it can be solved analytically, but that is not its only pleasant feature.
Like most good toy models, it serves as a nice prototype for what happens in
more complicated flows.  It also serves as a building block for what may be
called the \emph{local theory} of mixing (Section~\ref{sec:rsm}).

The perfect setting to consider a linear flow is in the limit of large Schmidt
number.  The Schmidt number is a dimensionless quantity defined as
\begin{equation*}
  \Sc \ldef \visc/\Diff
\end{equation*}
where~$\visc$ is the kinematic viscosity of the fluid and~$\Diff$ is the
diffusivity of the scalar.  The Schmidt number may be thought of as the ratio
of the diffusion time for the scalar to that for momentum in the fluid.
Alternatively, it can be regarded as the ratio of the (squared) length of the
smallest feature in the velocity field to that in the scalar field.  This last
interpretation is due to the fact that if~$\theta$ varies in space more
quickly than~$\sqrt{\Diff}$, then its gradient is large and diffusion wipes
out the variation.  The same applies to variations in the velocity field with
respect to~$\visc$.  Hence, for large Schmidt number the scalar field has much
faster variations than the velocity field.  This means that it is possible to
focus on a region of the domain large enough for the scalar concentration to
vary appreciably, but small enough that the velocity field appears linear.
Because there are many cases for which~$\Sc$ is quite large, this motivates
the use of a linear velocity field.  In fact, large~$\Sc$ number is the
natural setting for chaotic advection.  It is also the regime that was studied
by Batchelor and leads to the celebrated Batchelor
spectrum~\cite{Batchelor1959}. The limit of small~$\Sc$ is the domain of
\emph{homogenization theory} and of turbulent diffusivity models.  We shall
not discuss such things here.

\subsection{Solution of the Problem}

We choose a linear velocity field of the form
\begin{equation*}
  \velv = \xv\cdot\sigma(t), \qquad \Tr\sigma = 0,
\end{equation*}
where~$\sigma$ is a traceless matrix because~$\div\velv$ must vanish.
Inserting this into~\eqref{eq:adeq}, we want to solve the initial value
problem
\begin{equation}
  \pd_\time\theta + \xv\cdot\sigma(t)\cdot\grad\theta
  = \Diff\lapl\theta, \qquad \theta(\xv,0) = \theta_0(\xv).
  \label{eq:adeqlinv}
\end{equation}
Here the coordinate~$\xv$ is really a deviation from a reference fluid
trajectory.  (In Appendix~\ref{sec:adeqcomov} we derive~\eqref{eq:adeqlinv}
from~\eqref{eq:adeq} by transforming to a comoving frame and assuming the
velocity field is smooth.)  We will follow closely the solution of
Zeldovich~\etal~\cite{Zeldovich1984}, who solved this by the method of
``partial solutions.''  Consider a solution of the form
\begin{equation}
  \theta(\xv,\time) = \hat\theta(\kzv,\time)\exp(\imi\kv(\time)\cdot\xv),
  \qquad \kv(0) = \kzv,\quad
  \hat\theta(\kzv,0) = \hat\theta_0(\kzv),
  \label{eq:psol}
\end{equation}
where~$\kzv$ is some initial wavevector.
We will see if we can make this into a solution by a judicious choice
of~$\hat\theta(\kzv,\time)$ and~$\kv(\time)$.  The time derivative
of~\eqref{eq:psol} is
\begin{equation*}
  \pd_\time\theta =
  (\pd_\time\hat\theta + \imi\,\pd_\time\kv\cdot\xv\,\hat\theta)
  \exp(\imi\kv(\time)\cdot\xv)
\end{equation*}
and we have
\begin{equation*}
  \velv\cdot\grad\theta = \imi\,(\xv\cdot\sigma\cdot\kv)\,\hat\theta
  \exp(\imi\kv(\time)\cdot\xv).
\end{equation*}
Putting these together into~\eqref{eq:adeqlinv} and cancelling out the
exponential gives
\begin{equation*}
  \pd_\time\hat\theta + \imi\,\xv\cdot(\pd_\time\kv
  + \sigma\cdot\kv)\,\hat\theta = -\Diff\,\kc^2\hat\theta.
\end{equation*}
This must hold for all~$\xv$, and neither~$\hat\theta$ nor~$\kv$ depend
on~$\xv$, so we equate powers of~$\xv$.  This gives the two evolution
equations
\begin{subequations}
\begin{align}
  \pd_\time\kv &= -\sigma\cdot\kv\,,
  \label{eq:k}\\
  \pd_\time\hat\theta &= -\Diff\,\kc^2\hat\theta.
  \label{eq:thetah}
\end{align}
\label{eq:ktheta}
\end{subequations}
We can write the solution to~\eqref{eq:k} in terms of the \emph{fundamental
solution}~$\Tf(\time,0)$ as
\begin{equation*}
  \kv(\time) = \Tf(\time,0)\cdot\kzv\,,
\end{equation*}
where
\begin{equation}
  \pd_\time\Tf(\time,0) = -\sigma(\time)\cdot\Tf(\time,0),
  \qquad \Tf(0,0) = \Id
  \label{eq:Tfevol}
\end{equation}
and~$\Id$ is the identity matrix.  The advantage of doing this is that we can
use the same fundamental solution for all initial conditions.  We will usually
write~$\Tf_\time$ to mean~$\Tf(\time,0)$.  Note that because~$\Tr\sigma=0$, we
have
\begin{equation}
  \det\Tf_\time = 1.
  \label{eq:detT1}
\end{equation}
This is a standard result that is proved in Appendix~\ref{sec:detT1} for
completeness.  If~$\sigma$ is not a function of time, then the fundamental
solution is simply a matrix exponential,
\begin{equation*}
  \Tf_\time = \exp(-\sigma\,\time),
\end{equation*}
but in general the form of~$\Tf_\time$ is more complicated.

Now that we know the time-dependence of~$\kv$, we can express the solution
to~\eqref{eq:ktheta} as
\begin{subequations}
\begin{align}
  \kv(\time) &= \Tf_\time\cdot\kzv\,,\\
  \hat\theta(\kzv,\time) &= \hat\theta_0(\kzv)
  \exp\l\{-\Diff\int_0^\time\bigl(\Tf_\stime\cdot\kzv\bigr)^2\dint\stime\r\}.
  \label{eq:thetahatsol}
\end{align}
\end{subequations}
We can think of~$\Tf_\time$ as transforming a Lagrangian wavevector~$\kzv$ to
its Eulerian counterpart~$\kv$.  Thus~\eqref{eq:thetahatsol} expresses the
fact that~$\hat\theta$ decays diffusively at a rate determined by the
cumulative norm of the wavenumber~$\kv$ experienced during its evolution.

The full solution to~\eqref{eq:adeqlinv} is now given by superposition of the
partial solutions,
\begin{align}
  \theta(\xv,\time) &= \int\hat\theta(\kzv,\time)
  \exp(\imi\kv(\time)\cdot\xv)\dint^3\kzc\nonumber\\
  &= \int\hat\theta_0(\kzv)
  \exp\l\{\imi\,\xv\cdot\Tf_\time\cdot\kzv
  - \Diff\int_0^\time\bigl(\Tf_\stime\cdot\kzv\bigr)^2\dint\stime\r\}
  \!\dint^3\kzc\,,
  \label{eq:thetaxt}
\end{align}
where~$\hat\theta_0(\kzv)$ is the Fourier transform of the initial
condition~$\theta_0(\xv)$.%
\footnote{We are using the convention
\[
  \hat\theta(\kv) = \frac{1}{(2\pi)^{\sdim}}
  \int\theta(\xv)\,\ee^{-\imi\kv\cdot\xv}
  \dint^\sdim\xc\,,
\]
\[
  \theta(\xv) = \int\hat\theta(\kv)\,\ee^{\imi\kv\cdot\xv}
  \dint^\sdim\kc\,,
\]
for the Fourier transform in~$\sdim$ dimensions.
}
Assuming the spatial mean of~$\theta$ vanishes,
the variance~\eqref{eq:Var} is
\begin{equation*}
  \Var = \int\theta^2(\xv,\time)\dint^3\xc
  = \int\lvert\hat\theta(\kzv,\time)\rvert^2\dint^3\kzc\,,
\end{equation*}
which from~\eqref{eq:thetahatsol} becomes
\begin{equation}
  \Var = 
  \int\lvert\hat\theta_0(\kzv)\rvert^2
  \exp\l\{-2\Diff\int_0^\time\bigl(\Tf_\stime\cdot\kzv\bigr)^2\dint\stime\r\}
  \!\dint^3\kzc\,.
  \label{eq:Varlin}
\end{equation}
We thus have a full solution of the \adeq\ for the case of a linear velocity
field and found the time-evolution of the variance.  But what can be gleaned
from it?  We shall look at some special cases in the following section.

\subsection{Straining Flow in 2D}
\label{sec:strain}

We now take an even more idealised approach: consider the case where the
velocity gradient matrix $\sigma$ is constant.  Furthermore, let us restrict
ourselves to two-dimensional flows.  After a coordinate change, the traceless
matrix~$\sigma$ can only take two possible forms,
\begin{equation}
  \sigma^{\text{(2a)}} =
  \begin{pmatrix}\lambda & 0 \\ 0 & -\lambda\end{pmatrix}
  \qquad \text{and} \qquad
  \sigma^{\text{(2b)}} = \begin{pmatrix}0 & 0 \\ U' & 0\end{pmatrix}.
    \label{eq:sigma2D}
\end{equation}
Case (2a) is a purely straining flow that stretches exponentially in one
direction, and contracts in the other.  Case (2b) is a linear shear flow in
the~$\xc_1$ direction.  We assume without loss of generality that~$\lambda>0$
and~$U'>0$.  The form~$\sigma^{(2b)}$ is known as the Jordan canonical form,
and can only occur for degenerate eigenvalues.  Since by incompressibility the
sum of these identical eigenvalues must vanish, they must both vanish.  The
corresponding fundamental matrices~\hbox{$\Tf_\time = \exp(-\sigma\,\time)$}
are
\begin{equation}
  \Tf_\time^{\text{(2a)}} =
  \begin{pmatrix}\ee^{-\lambda\time} & 0 \\
    0 & \ee^{\lambda\time}\end{pmatrix}
  \qquad \text{and} \qquad
  \Tf_\time^{\text{(2b)}} =
  \begin{pmatrix}1 & 0 \\ -U'\time & 1\end{pmatrix}.
    \label{eq:Tf2D}
\end{equation}
These are easy to compute: in the first instance one merely exponentiates the
diagonal elements, in the second the exponential power series terminates after
two terms, because the square of~$\sigma^{(2b)}$ is zero.

Let us consider Case (2a), a flow with constant stretching (the case
considered by Batchelor~\cite{Batchelor1959}).  The action of the fundamental
matrix on~$\kv_0$ for Case (2a) is
\begin{equation}
  \Tf^{(2a)}_\time\cdot\kzv = \l(\ee^{-\lambda\time}\,{\kzc}_1\,,\,
  \ee^{\lambda\time}\,{\kzc}_2\r),
  \label{eq:Tf2akzv}
\end{equation}
with norm
\begin{equation}
  \bigl(\Tf^{(2a)}_\time\cdot\kzv\bigr)^2 = \ee^{-2\lambda\time}\,{\kzc}_1^2
  + \ee^{2\lambda\time}\,{\kzc}_2^2\,.
  \label{eq:k2anorm}
\end{equation}
The wavevector~$\kv(\time) = \Tf^{(2a)}_\time\cdot\kzv$ grows exponentially in
time, which means that the length scale is becoming very small.  This only
occurs in the direction~$\xc_2$, which is sensible because that direction
corresponds to a contracting flow.  Picture a curtain being closed: the
bunching up of the fabric into tight folds is analogous to the contraction.
(Of course, it is difficult to close a curtain exponentially quickly forever!)
The component of the wavevector in the~$\xc_1$ direction decreases in
magnitude, which corresponds to the opening of a curtain.

Let's see what happens to one Fourier mode.  By inserting~\eqref{eq:k2anorm}
in~\eqref{eq:thetahatsol}, we have
\begin{equation*}
  \hat\theta(\kzv,\time) = \hat\theta_0(\kzv)
  \exp\l\{-\Diff\int_0^\time\bigl(\ee^{-2\lambda\stime}\,{\kzc}_1^2
  + \ee^{2\lambda\stime}\,{\kzc}_2^2\bigr)\dint\stime\r\}\,.
\end{equation*}
The time-integral can be done explicitly, and we find
\begin{equation*}
  \hat\theta(\kzv,\time)
  = \hat\theta_0(\kzv) \exp\l\{-\frac{\Diff}{2\lambda}\l(
  \l(\ee^{2\lambda\time}-1\r){\kzc}_2^2
  - \l(\ee^{-2\lambda\time}-1\r){\kzc}_1^2\r)\r\}\,.
\end{equation*}
For moderately long times ($\time\gtrsim\lambda^{-1}$), we can surely
neglect~$\ee^{-2\lambda\time}$ compared to~$1$, and~$1$ compared
to~$\ee^{2\lambda\time}$,
\begin{equation}
  \hat\theta(\kzv,\time) \simeq \hat\theta_0(\kzv)
  \exp\l\{-\frac{\Diff}{2\lambda}\l(\ee^{2\lambda\time}\,{\kzc}_2^2
  + {\kzc}_1^2\r)\r\}\,.
  \label{eq:Varsigconst}
\end{equation}
Actually, this assumption of moderately long time is easily justified
physically.  If~\hbox{$\Diff\kc^2/\lambda \ll 1$}, where~$\kc$ is the
largest initial wavenumber (that is, the smallest initial scale), then the
argument of the exponential in~\eqref{eq:Varsigconst} is small, unless
\begin{equation}
  \ee^{2\lambda\time} \gtrsim \Pe
  \label{eq:egPe}
\end{equation}
where the \emph{P\'eclet number} is
\begin{equation}
  \Pe = \frac{\lambda}{\Diff\,\kc^2}\,.
  \label{eq:Peus}
\end{equation}
Thus the assumption that~$\ee^{2\lambda\time}$ is large is a consequence
of~$\Pe$ being large, since otherwise the exponential
in~\eqref{eq:Varsigconst} is near unity and can be ignored---variance is
approximately constant.  We can turn~\eqref{eq:egPe} into a requirement on
the time,
\begin{equation}
  \lambda\,\time \gtrsim \log\Pe^{1/2}\,.
  \label{eq:tglogPe}
\end{equation}
It is clear from~\eqref{eq:tglogPe} that~$\lambda^{-1}$ sets the time scale
for the argument of the exponential in~\eqref{eq:Varsigconst} to become
important.  The P\'eclet number influences this time scale only weakly
(logarithmically).  This is probably the most important physical fact about
chaotic mixing: \emph{Small diffusivity has only a logarithmic effect}.  Thus
vigorous stirring always has a chance to overcome a small diffusivity, no
matter how small: we need just stir a bit longer.

Note that the variance is given by
\begin{equation*}
  \Var = \int\lvert\hat\theta_0(\kzv)\rvert^2
  \exp\l\{-\frac{\Diff}{\lambda}\l(\ee^{2\lambda\time}\,{\kzc}_2^2
  + {\kzc}_1^2\r)\r\}\!\dint^2\kzc\,,
\end{equation*}
which is approximately constant for~$\time\ll\lambda^{-1}\log\Pe^{1/2}$.
This does \emph{not} mean that the concentration field
\begin{equation}
  \theta(\xv,\time) = \int\hat\theta(\kzv,\time)\,\ee^{\imi\kv(\time)\cdot\xv}
  \dint^2\kzc
  \label{eq:thetaxt2}
\end{equation}
is constant, even if~$\hat\theta(\kzv,\time)$ is,
because~\hbox{$\kv(\time)=\Tf^{(2a)}_\time\cdot\kzv$} is a function of time
from~\eqref{eq:Tf2akzv}.  This time dependence becomes important
for~$\time\gtrsim\lambda^{-1}$.

The P\'eclet number may be thought of as the ratio of the advection time of
the flow to the diffusion time for the scalar. It is usually written as
\begin{equation}
  \Pe \ldef UL/\Diff\,,
  \label{eq:Pe}
\end{equation}
where~$U$ is a typical velocity and~$L$ a typical length scale.  Our velocity
estimate in~\eqref{eq:Peus} is~$\lambda/\kc$, and our length scale is~$\kc$,
which are both natural for the problem at hand.  Just like large~$\Sc$,
large~$\Pe$ is the natural setting for chaotic advection.  In fact, if~$\Pe$
is small then diffusion is faster than advection, and stirring is not really
required!  Large~$\Pe$ means that diffusion by itself is not very effective,
so that stirring is required.  We shall always assume that~$\Pe$ is large.

We return to~\eqref{eq:Varsigconst}: the striking thing about that equation is
its prediction for the rate of decay of the concentration field.  Roughly
speaking,~\eqref{eq:Varsigconst} predicts
\begin{equation}
  \theta(\xv,\time) \sim \exp\l\{-\Pe^{-1}\,\ee^{2\lambda\time}\r\}
  \label{eq:Varsuper}
\end{equation}
for~\hbox{$\lambda\time\gg 1$}.  Eq.~\eqref{eq:Varsuper} is the exponential of
an exponential---a \emph{superexponential} decay.  This is \emph{extremely}
fast decay.  In fact, unnaturally so: it is hard to imagine a physically
sensible system that could mix this quickly.  Something more has to be at work
here.

If we examine~\eqref{eq:Varsigconst} closely, we see that the culprit is the
term
\begin{equation}
  \ee^{2\lambda\time}\,{\kzc}_2^2\,,
  \label{eq:culprit}
\end{equation}
which grows exponentially fast.  This term has its origin in the Laplacian in
the \adeq~\eqref{eq:adeq}: the contracting direction of the flow (the~$\xc_2$
direction) leads to an exponential increase in the wavenumber via the
curtain-closing mechanism.  This is exactly the mechanism for enhanced mixing
we advertised on p.~\pageref{page:thecatch}: very large gradients of
concentration are being created, exponentially fast.  This mechanism is just
acting too quickly for our taste!

So what's the problem?  We are doing the wrong thing to obtain our
estimate~\eqref{eq:Varsuper}.  This estimate tells us how fast a typical
wavevector decays, and it says that this occurs very quickly.  What we really
want to know is what modes survive superexponential decay the longest, and at
what rate \emph{they} decay.  Clearly the concentration in most wavenumbers
gets annihilated almost instantly, once the condition~\eqref{eq:tglogPe} is
satisfied.  But a small number remains: those are the modes with wavevector
closely aligned to the~$\xc_1$ (stretching) direction, or equivalently that
have a very small projection on the~$\xc_2$ (contracting) direction.  To
overcome the exponential growth in~\eqref{eq:culprit}, we require
\begin{equation}
  {\kzc}_2 \sim \ee^{-\lambda\time}\,,
  \label{eq:aligned}
\end{equation}
that is at any given time we need consider only wavenumbers
satisfying~\eqref{eq:aligned}, since the concentration in all the others has
long since been wiped out by diffusion.  The consequence is that
the~${\kzc}_2$ integral in~\eqref{eq:thetaxt2} is dominated by these surviving
modes.  To see this, we blow up the~${\kzc}_2$ integration by making the
coordinate change~${\kzct}_2 = {\kzc}_2\,\ee^{\lambda\time}$
in~\eqref{eq:thetaxt2},
\begin{multline}
  \theta(\xv,\time) = \ee^{-\lambda\time}\int_{-\infty}^\infty\dint{\kzc}_1
  \int_{-\infty}^{\infty}
  \dint{\kzct}_2\,\,\hat\theta_0({\kzc}_1,{\kzct}_2\,\ee^{-\lambda\time})\\
  \times\,\ee^{\imi\kv(\time)\cdot\xv}
  \exp\l\{-\frac{\Diff}{2\lambda}\l({\kzct}_2^2 + {\kzc}_1^2\r)\r\},
  \label{eq:Varsigconst2}
\end{multline}
The decay factor~$\ee^{-\lambda\time}$ has appeared in front.
For small diffusivity, we can neglect the~${\kzc}_1^2$ term in the
exponential (it just smooths out the initial concentration field a little).%
\footnote{We require the initial condition to be smooth at small scales.
Here's why: for small $\Diff$, ${\kzc}_1$ needs to be large to matter in the
argument of the exponential.  But a smooth $\theta$ decays exponentially with
${\kzc}_1$, so there is no variance in these modes anyways.}
We can then take the inverse Fourier transform
of~$\hat\theta_0({\kzc}_1,{\kzct}_2\,\ee^{-\lambda\time})$,
\begin{equation*}
  \hat\theta_0({\kzc}_1,{\kzct}_2\,\ee^{-\lambda\time})
  = \frac{1}{(2\pi)^2} \int\theta_0(\tilde\xv)\,
  \exp\l(-\imi{\kzc}_1\tilde\xc_1
  - \imi{\kzct}_2\,\ee^{-\lambda\time}\tilde\xc_2\r)
  \dint\tilde\xc_1\dint\tilde\xc_2
\end{equation*}
and insert this into~\eqref{eq:Varsigconst2}.  We then interchange the order
of integration: the~${\kzc}_1$ integral gives a~$\delta$-function, and
the~${\kzct}_2$ integral gives a Gaussian.  The final result is
\begin{equation}
  \theta(\xv,\time) = \ee^{-\lambda\time}
  \int_{-\infty}^\infty
  \,\,\theta_0(\ee^{-\lambda\time}\xc_1,\tilde\xc_2)
  \,\Gauss\bigl(\xc_2 - \ee^{-\lambda\time}\,\tilde\xc_2\,;\,\fwid\bigr)
  \dint\tilde\xc_2\,,
  \label{eq:thetastabilised2}
\end{equation}
where
\begin{equation}
  \Gauss(\xc;\fwid) \ldef \frac{1}{\sqrt{2\pi\fwid^2}}\,\ee^{-x^2/2\fwid^2}
  \label{eq:Gaussdef}
\end{equation}
is a normalised Gaussian distribution with standard deviation~$\fwid$, and
we defined the length
\begin{equation*}
  \fwid \ldef \sqrt{{\Diff}/{\lambda}}\,.
\end{equation*}
If the initial concentration decays for large~$\lvert\xc_2\rvert$ (as when we
have a single blob of dye), then~\eqref{eq:thetastabilised2} can be simplified
to
\begin{equation}
  \theta(\xv,\time) = \ee^{-\lambda\time}\,
  \Gauss\bigl(\xc_2\,;\,\fwid\bigr)\int_{-\infty}^\infty
  \,\,\theta_0(\ee^{-\lambda\time}\xc_1,\tilde\xc_2)
  \dint\tilde\xc_2\,.
  \label{eq:thetastabilised3}
\end{equation}

So the~$\xc_1$ dependence in~\eqref{eq:thetastabilised3} is given by the
``stretched'' initial distribution, averaged over~$\xc_2$.  The important
thing to notice is that
\begin{equation}
  \theta(\xv,\time) \sim \ee^{-\lambda\time}\,.
  \label{eq:Varexp}
\end{equation}
This is a much more reasonable estimate for the decay of concentration
than~\eqref{eq:Varsuper}!  The concentration thus decays exponentially at a
rate given by the rate-of-strain (or stretching rate) in our flow.  The
exponential decay is entirely due to the narrowing of the domain for eligible
(\ie, nondecayed) modes.  This ``domain of eligibility'' is also known as the
\emph{cone} or the \emph{cone of safety}~\cite{Zeldovich1984,Thiffeault2004b}.
(In two dimensions it is more properly called a wedge.)  The concentration
associated with wavevectors that fit within this cone is temporarily shielded
from being diffusively wiped out, but as the aperture of the cone is shrinking
exponentially more and more modes leave the safety of the cone as time
progresses.  \comment{Figure}

Notice that~\eqref{eq:Varexp} is independent of~$\Diff$.  This brings us to
the second most important physical fact about chaotic mixing: \emph{The
asymptotic decay rate of the concentration field tends to be independent of
diffusivity}.  But note that a nonzero diffusivity is crucial in forcing the
alignment~\eqref{eq:aligned}.  The only effect of the diffusivity is to
lengthen the wait before exponential decay sets in, as given by the
estimate~\eqref{eq:tglogPe}.  But this effect is only logarithmic in the
diffusivity.

\begin{figure}
\centering
\includegraphics[width=\textwidth]{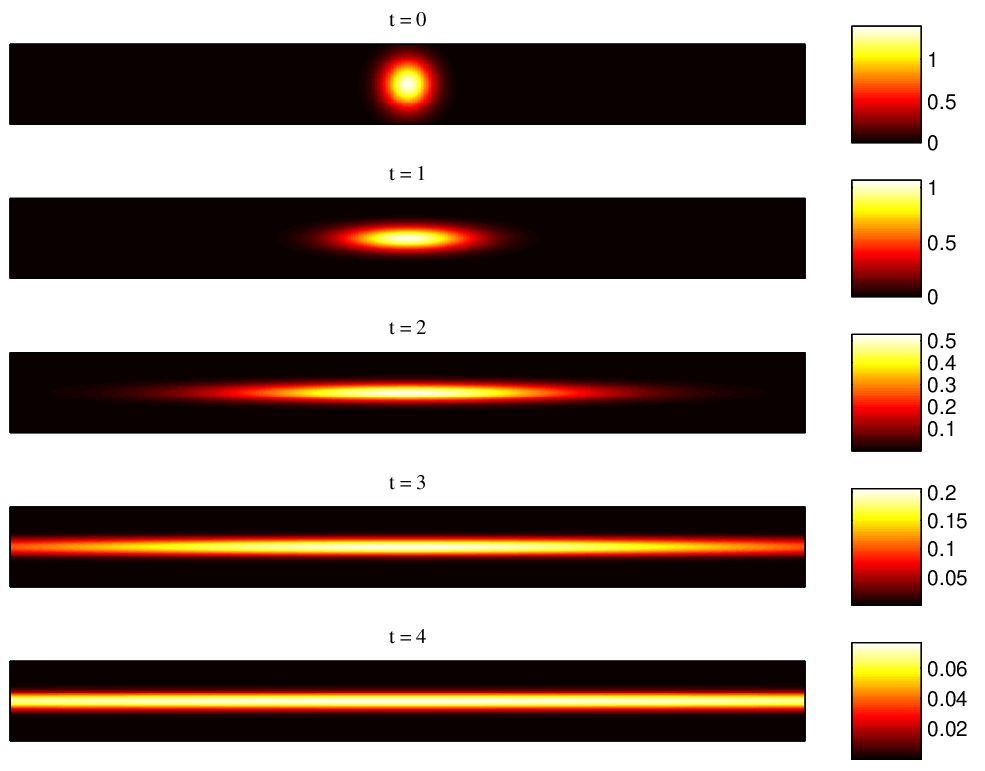}
\caption{A patch of dye in a uniform straining flow.  The amplitude of the
  concentration field decreases exponentially with time.  The length of the
  filament increases exponentially, whilst its width is stabilised
  at~$\fwid=\sqrt{\Diff/\lambda}$.}
\label{fig:strainblob}
\end{figure}

We can also try to think of~\eqref{eq:thetastabilised3} in real rather than
Fourier space.  Consider an initial distribution of concentration.  Our
straining flow will stretch this distribution in the~$\xc_1$ direction, and
contract it in the~$\xc_2$ direction.  Gradients in~$\xc_2$ will thus become
very large, so that eventually diffusion will limit further contraction in
the~$\xc_2$ direction and the distribution will stabilise with
width~$\sqrt{\Diff/\lambda}$ (see Fig.~\ref{fig:strainblob}).  This is what
the Gaussian prefactor in~\eqref{eq:thetastabilised3} is telling us: the
asymptotic distribution has ``forgotten'' its initial shape in~$\xc_2$.  We
say that the contracting direction has been \emph{stabilised}.

\subsection{Shear Dispersion in 2D}
\label{sec:shear}

So far we have only considered case (2a) in~\eqref{eq:sigma2D}.  For case
(2b), we have from~\eqref{eq:Tf2D}
\begin{equation*}
  \Tf^{(2b)}_\time\cdot\kzv = \l({\kzc}_1,
  {\kzc}_2 - U'\time\,{\kzc}_1\r)
\end{equation*}
with norm
\begin{equation}
  \bigl(\Tf^{(2b)}_\time\cdot\kzv\bigr)^2
  = {\kzc}_1^2 + \l({\kzc}_2 - U'\time\,{\kzc}_1\r)^2\,.
  \label{eq:k2bnorm}
\end{equation}
Inserting~\eqref{eq:k2bnorm} in~\eqref{eq:thetaxt}, we have
\begin{equation*}
  \theta(\xv,\time) = \int\hat\theta_0(\kzv)\,\ee^{\imi\kv(\time)\cdot\xv}
  \exp\l\{-\Diff\int_0^\time
  \bigl({\kzc}_1^2 + \l({\kzc}_2 - U'\stime\,{\kzc}_1\r)^2
    \bigr)\dint\stime\r\}
  \!\dint^2\kzc\,.
\end{equation*}
We can then explicitly do the time integral in the exponential,
\begin{multline}
  \theta(\xv,\time) = \int\hat\theta_0(\kzv)\,\ee^{\imi\kv(\time)\cdot\xv}\\
  \times\exp\l\{-\Diff\,{\kzc}_1^2\,\time
  - \frac{\Diff}{3U'{\kzc}_1}\,
  \bigl(\l(U'\time\,{\kzc}_1 - {\kzc}_2\r)^3
  + {\kzc}_2^3\bigr)
  \r\}\!\dint^2\kzc\,.
  \label{eq:Varsheardisp}
\end{multline}
The enhancement to diffusion in this case is reflected in the cubic power of
time in the exponential.  This is not as strong as the exponential enhancement
of case (2a), but is nevertheless very significant.  This phenomenon is known
as \emph{shear dispersion} or \emph{Taylor dispersion}.  The mechanism is
often called the \emph{venetian blind} effect.  Assuming the initial
distribution~$\theta_0$ depends only on~${\kzc}_1$, then lines of constant
concentration which are initially vertical are tilted by the shear flow, in a
manner reminiscent of venetian blinds.  The distance between the lines of
constant concentration decreases with time as~$(U'\time)^{-1}$, which gives an
effective enhancement to diffusion.  The time required to overcome a weak
diffusivity is thus
\begin{equation}
  U'\time \gtrsim ({\kzc}_1^2\,\Diff/U')^{-1/3}.
  \label{eq:tgk3}
\end{equation}
If we use~${\kzc}_1^{-1}$ as
a length scale and~$U'$ as a time scale, we can define a P\'eclet number $\Pe
\ldef U'/({\kzc}_1{}^2\,\Diff)$ and rewrite~\eqref{eq:tgk3} as
\begin{equation}
  U'\time \gtrsim \Pe^{1/3}
  \label{eq:tgPe3}
\end{equation}
which should be compared to~\eqref{eq:tglogPe}, the corresponding expression
for the case (2a).  Here there is a power law dependence on the P\'eclet
number, rather than logarithmic, so we may have to wait a long time for
diffusion to become important.  This makes the linear velocity field
approximation more likely to break down.

Let us consider the time-asymptotic limit~$U'\time \gg 1$: we might then be
tempted to neglect everything but the~$U'\time\,{\kzc}_1$ term in the argument
of the exponential in~\eqref{eq:Varsheardisp}.  However, this would be a
mistake.  To see more clearly what happens, define the dimensionless
time~\hbox{$\tau \ldef U'\time$} and the length~\hbox{$\shrlen^2=\Diff/U'$}.
Equation~\eqref{eq:Varsheardisp} then becomes
\begin{multline}
  \theta(\xv,\time) = \int\hat\theta_0(\kzv)\,\ee^{\imi\kv(\time)\cdot\xv}\\
  \times\exp\l\{-\shrlen^2\l({\kzc}_1^2\tau + {\kzc}_2^2\tau
  + \tfrac{1}{3}{\kzc}_1^2\tau^3 - {\kzc}_1{\kzc}_2\tau^2\r)
  \r\}\!\dint^2\kzc\,.
  \label{eq:sheardisp2}
\end{multline}
The first two terms in the exponential are just what is expected of regular
diffusion in the absence of flow.  The next term is the enhancement to
diffusion along the~$\xc_1$ direction: it will force the
modes~${\kzc}_1\sim\tau^{-3/2}$ to be dominant, since everything else will be
damped away.  Similarly, the last term forces~${\kzc}_2\sim\tau^{-1/2}$.
Assuming these scalings, the only term that can be neglected
for~\hbox{$\tau\gg 1$} is the very first one,~${\kzc}_1^2\tau$.

\begin{figure}
\centering
\includegraphics[width=\textwidth]{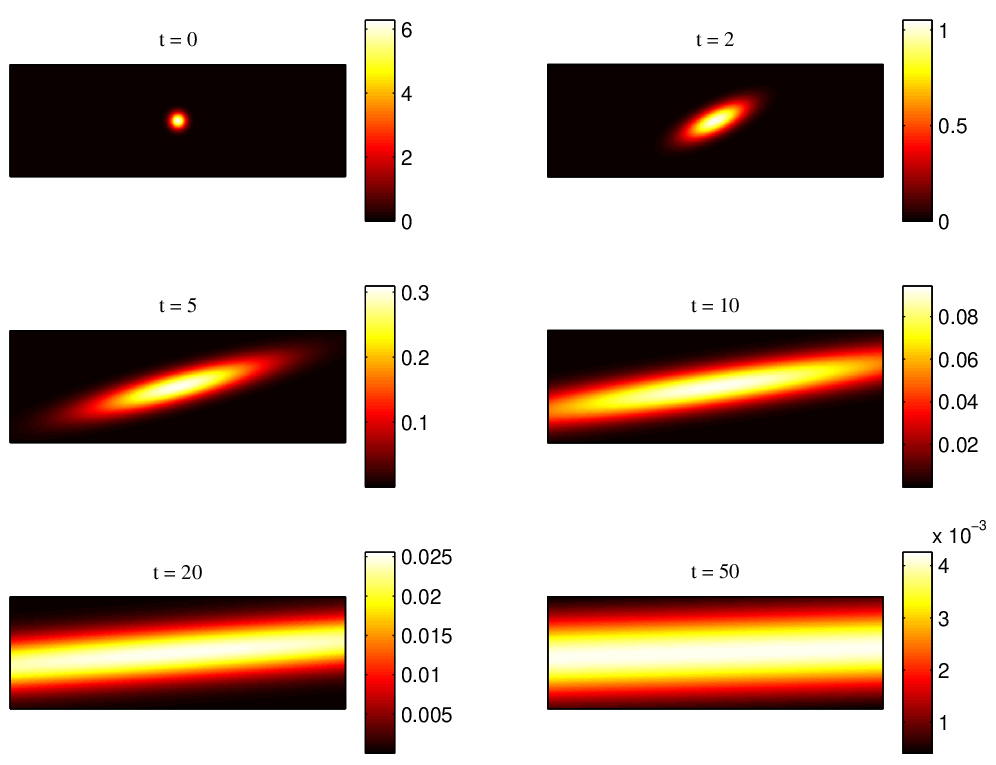}
\caption{A patch of dye in a uniform shearing flow.  The amplitude of the
  concentration field decreases algebraically with time as~$\time^{-2}$.  The
  length of the filament increases as~$\time^{3/2}$, whilst its width
  increases as~$\time^{1/2}$.}
\label{fig:shearblob}
\end{figure}

We make the change of
variable~${\kzct}_1=\tau^{3/2}{\kzc}_1$,~${\kzct}_2=\tau^{1/2}{\kzc}_2$
in~\eqref{eq:sheardisp2},
\begin{multline}
  \theta(\xv,\time) = \tau^{-2}\int
  \hat\theta_0({\kzct}_1\,\tau^{-3/2}\,,\,
  {\kzct}_2\,\tau^{-1/2})\,
  \ee^{\imi(\tau^{-3/2}\xc_1-\tau^{-1/2}\xc_2){\kzct}_1
    + \imi\tau^{-1/2}\xc_2 {\kzct}_2}\\
  \times\exp\l\{-\shrlen^2 \l({\kzct}_2^2 + \tfrac{1}{3}{\kzct}_1^2
  - {\kzct}_1 {\kzct}_2\r)\r\}\!\dint{\kzct}_1\dint {\kzct}_2\,.
  \label{eq:sheardisp3}
\end{multline}
If we approximate~$\hat\theta_0({\kzct}_1\,\tau^{-3/2}\,,\,
{\kzct}_2\,\tau^{-1/2}) \simeq \hat\theta_0(0,0)$, we can do the integrals
in~\eqref{eq:sheardisp3} and find
\begin{equation}
  \theta(\xv,\time) \simeq 2\sqrt{3}\pi\,\shrlen^{-2}\,\tau^{-2}\,
  \hat\theta_0(0,0)\,
  \exp\l\{-\frac{3\xc_1^2 - 3\xc_1\xc_2\tau+\xc_2^2\tau^2}
		{\shrlen^2\tau^3}\r\}.
  \label{eq:sheardisp4}
\end{equation}
For moderate values of~$\xc_1$ ($\xc_1\ll\shrlen\tau$), we have
\begin{equation}
  \theta(\xv,\time) \simeq 2\sqrt{3}\pi\,\shrlen^{-2}\,\tau^{-2}\,
  \ee^{-\xc_2^2/\shrlen^2\tau}\,
  \hat\theta_0(0,0).
  \label{eq:sheardisp5}
\end{equation}
The width in the~$\xc_2$ direction of an initial distribution thus increases
as~$\shrlen\tau^{1/2}=\sqrt{\Diff\,\time}$.  This is independent of~$U'$ and
is exactly the same as expected from pure diffusion.  The width in the~$\xc_1$
direction in~\eqref{eq:sheardisp4} increases
as~$\shrlen\tau^{3/2}=U'\time\sqrt{\Diff\,\time}$ (see
Fig.~\ref{fig:shearblob}).

\subsection{Three Dimensions}
\label{sec:3D}

In three dimensions, there are three basic forms for the matrix $\sigma$:
\begin{equation*}
  \sigma^{\text{(3a)}} =
  \begin{pmatrix}\lambda_1 & 0 & 0\\ 0 & \lambda_2 & 0\\
  0 & 0 & -\lambda_1-\lambda_2
  \end{pmatrix}
  \ \text{;} \quad
  \sigma^{\text{(3b)}} = \begin{pmatrix}0 & 0 & 0 \\
    U' & 0 & 0\\ 0 & U' & 0 \end{pmatrix}
  \ \text{;} \quad
  \sigma^{\text{(3c)}} = \begin{pmatrix}\lambda & 0 & 0 \\
    U' & \lambda & 0\\ 0 & 0 & -2\lambda \end{pmatrix},
\end{equation*}
with corresponding fundamental matrices
\begin{gather} \Tf^{\text{(3a)}} =
  \begin{pmatrix}\ee^{-\lambda_1\time} & 0 & 0\\ 0 & \ee^{-\lambda_2\time}
    & 0\\ 0 & 0 & \ee^{(\lambda_1+\lambda_2)\time}
  \end{pmatrix}
  \ \text{;} \quad
  \Tf^{\text{(3b)}} = \begin{pmatrix}0 & 0 & 0 \\
    -U'\time & 0 & 0\\ \tfrac{1}{2}(U'\time)^2 & -U'\time & 0 \end{pmatrix}
  \ \text{;}\\
  \Tf^{\text{(3c)}} = \begin{pmatrix}\ee^{-\lambda\time} & 0 & 0 \\
    -U'\time\,\ee^{-\lambda\time} & \ee^{-\lambda\time} & 0\\
    0 & 0 & \ee^{2\lambda\time} \end{pmatrix}.
\end{gather}
We can assume without loss of generality that~$\lambda_1\ge0$,
$\lambda_1\ge\lambda_2$, and $U'>0$, but the sign of~$\lambda_2$ and~$\lambda$
is arbitrary; however, we must
have \hbox{$\lambda_3=-\lambda_1-\lambda_2\le0$}.  The case of greatest
interest to us is (3a).  The relevant~$\kv(\time)$, corresponding
to~\eqref{eq:Tf2akzv}, is
\begin{equation}
  \Tf^{(3a)}_\time\cdot\kzv = \l(\ee^{-\lambda_1\time}\,{\kzc}_1 \,,\,
  \ee^{-\lambda_2\time}\,{\kzc}_2 \,,\,
  \ee^{\lvert\lambda_3\rvert\time}\,{\kzc}_3\r)\,,
  \label{eq:Tf3akzv}
\end{equation}
which is used in~\eqref{eq:thetaxt} to give
\begin{multline}
  \theta(\xv,\time) = \int\hat\theta_0(\kzv)\,\ee^{\imi\kv(\time)\cdot\xv}
  \exp\Bigl\{-\tfrac{1}{2}\Diff\Bigl(
  \lambda_1^{-1}\l(1 - \ee^{-2\lambda_1\time}\r){\kzc}_1^2\\
  + \lambda_2^{-1}\l(1 - \ee^{-2\lambda_2\time}\r){\kzc}_2^2
  + \lvert\lambda_3\rvert^{-1}
  \l(\ee^{2\lvert\lambda_3\rvert\time}-1\r){\kzc}_3^2\Bigr)
  \Bigr\}\dint^3\kzc\,.
\end{multline}
What happens next depends on the sign of~$\lambda_2$: the question is
whether~$\ee^{-2\lambda_2\time}$ grows or decays
for~$\time\gg\lvert\lambda_2\rvert^{-1}$.  If~$\lambda_2>0$, then we have
\begin{multline}
  \theta(\xv,\time) \simeq \int\hat\theta_0(\kzv)\,\ee^{\imi\kv(\time)\cdot\xv}
  \exp\Bigl\{-\tfrac{1}{2}\Diff\Bigl(
  \lambda_1^{-1}{\kzc}_1^2\\ + \lambda_2^{-1}{\kzc}_2^2
  + \lvert\lambda_3\rvert^{-1}\,
  \ee^{2\lvert\lambda_3\rvert\time}\,{\kzc}_3^2\Bigr)
  \Bigr\}\dint^3\kzc\,,
  \label{eq:l2gt0}
\end{multline}
whilst for~$\lambda_2<0$
\begin{multline}
  \theta(\xv,\time) \simeq \int\hat\theta_0(\kzv)\,\ee^{\imi\kv(\time)\cdot\xv}
  \exp\Bigl\{-\tfrac{1}{2}\Diff\Bigl(
  \lambda_1^{-1}{\kzc}_1^2\\
  + \lvert\lambda_2\rvert^{-1}\ee^{2\lvert\lambda_2\rvert\time}{\kzc}_2^2
  + \lvert\lambda_3\rvert^{-1}\,
  \ee^{2\lvert\lambda_3\rvert\time}{\kzc}_3^2\Bigr)
  \Bigr\}\dint^3\kzc\,.
  \label{eq:l2lt0}
\end{multline}
Both approximations are valid when~$\time \gg
\max(\lambda_1^{-1}\,,\,\lvert\lambda_2\rvert^{-1})$.  For~$\lambda_2=0$ the
situation is similar to the two-dimensional case (2a):
\begin{equation*}
  \theta(\xv,\time) \simeq \int\hat\theta_0(\kzv)\,\ee^{\imi\kv(\time)\cdot\xv}
  \exp\Bigl\{-\frac{\Diff}{2\lambda_1}\Bigl({\kzc}_1^2
  + \ee^{2\lambda_1\time}{\kzc}_3^2\Bigr)
  \Bigr\}\dint^3\kzc\,,
\end{equation*}
valid when~\hbox{$\time \gg \lambda_1^{-1}$}.

The rest of the calculation is very similar to the two-dimensional case (2a),
in going from~\eqref{eq:thetaxt2} to~\eqref{eq:thetastabilised3}.  In
both~\eqref{eq:l2gt0} and~\eqref{eq:l2lt0} the~$\xc_3$ direction is
stabilised, that is we need to blow up the~${\kzc}_3$ integral to remove the
time dependence from the exponential, and find that the integral is dominated
by~\hbox{${\kzc}_3\simeq 0$}.  The~$\xc_2$ direction is also stabilised
in~\eqref{eq:l2lt0}, so we can set~${\kzc}_2\simeq 0$.  We thus find
for~\hbox{$\lambda_2 \ge 0$},
\begin{equation}
  \theta(\xv,\time) \simeq
  \ee^{-\lvert\lambda_3\rvert\time}\,\Gauss\bigl(\xc_3;\fwid_3\bigr)
  \int\theta_0(\ee^{-\lambda_1\time}\xc_1,\ee^{-\lambda_2\time}\xc_2,
  \tilde\xc_3)\dint\tilde\xc_3\,,
  \label{eq:pancakes}
\end{equation}
and for~\hbox{$\lambda_2 < 0$},
\begin{equation}
  \theta(\xv,\time) \simeq
  \ee^{-(\lvert\lambda_2\rvert + \lvert\lambda_3\rvert)\time}\,
  \Gauss\bigl(\xc_2;\fwid_2\bigr)\,
  \Gauss\bigl(\xc_3;\fwid_3\bigr)\int
  \theta_0(\ee^{-\lambda_1\time}\xc_1,\tilde\xc_2,
  \tilde\xc_3)\dint\tilde\xc_2\dint\tilde\xc_3\,,
  \label{eq:ropes}
\end{equation}
where~\hbox{$\fwid_i \ldef \sqrt{\Diff/\lvert\lambda_i\rvert}$}.  Contracting
directions have their spatial dependence given by a time-independent Gaussian,
with an overall exponential decay; stretching directions do just that: they
stretch the initial distribution, with no diffusive effect.  Solutions of the
form~\eqref{eq:pancakes} are called \emph{pancakes}, and those of the
form~\eqref{eq:ropes} are called \emph{ropes} or \emph{tubes}.

There is another way of thinking about the asymptotic
forms~\eqref{eq:thetastabilised3},~\eqref{eq:pancakes},
and~\eqref{eq:ropes}~\cite{Balkovsky1999}: contracting directions are
stabilised near some constant width $\fwid_j$, and expanding directions lead
to exponential growth of the width of an initial distribution along the
direction.  Thus, the volume of the initial distribution grows exponentially
at a rate given by the sum of~$\lambda_i$'s associated with stretching
directions, but the total amount of~$\theta$ remains fixed (the mean is
conserved).  Hence, the concentration at a point should decay inversely
proportional to the volume, which is exactly
what~\eqref{eq:thetastabilised3},~\eqref{eq:pancakes}, and~\eqref{eq:ropes}
predict.

\section{Random Strain Models}
\label{sec:rsm}

In Section~\ref{sec:linear} we analysed the deformation of a patch of
concentration field (a `blob') in a linear velocity field.  Though this is
interesting in itself, it is a far cry from reality.  We will now inch
slightly closer to the real world by giving a random time dependence to our
velocity field.

\subsection{A Single Blob}
\label{sec:oneblob}

\begin{figure}
\psfrag{e1}{\hspace{-1em}$\ee^{-\lambda^{(1)}\tau}$}
\psfrag{e2}{\hspace{-1em}$\ee^{-\lambda^{(2)}\tau}$}
\psfrag{e3}{\hspace{-1em}$\ee^{-\lambda^{(3)}\tau}$}
\psfrag{e4}{\hspace{-1em}$\ee^{-\lambda^{(4)}\tau}$}
\psfrag{e5}{\hspace{-1em}$\ee^{-\lambda^{(5)}\tau}$}
\psfrag{e6}{\hspace{-1em}$\ee^{-\lambda^{(6)}\tau}$}
\centering
\includegraphics[width=.7\textwidth]{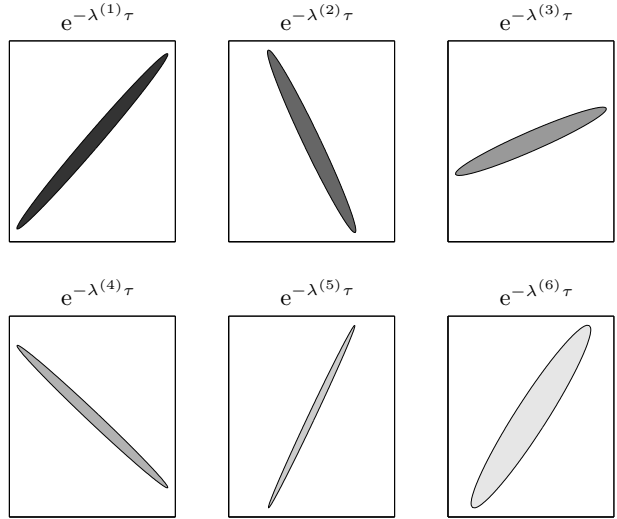}
\caption{A single blob being stretched for a time~$\tau$ by successive
random straining flows.  The amplitude of the concentration field decays
by~$\exp(-\lambda^{(i)}\tau)$ at each period.}
\label{fig:randomstrainblob}
\end{figure}

Consider a single blob in a two-dimensional linear velocity field of the type
we treated in Section~\ref{sec:strain} (case (2a)).  Now assume the orientation
and stretching rate~$\lambda$ of the straining flow change randomly every
time~$\tau$.  This situation is depicted schematically in
Figure~\ref{fig:randomstrainblob}.
We assume that the time~$\tau$ is much larger than a typical stretching
rate~$\lambda$, so that there is sufficient time for the blob to be deformed
into its asymptotic form~\eqref{eq:thetastabilised3} at each period, which
predicts that at each period the concentration field will decrease by a
factor~$\exp(-\lambda^{(i)}\tau)$, where~$\lambda^{(i)}$ is the stretching
rate at the~$i$th period.  The concentration field after~$n$ periods will thus
be proportional to the product of decay factors,
\begin{align}
  \theta &\sim \ee^{-\lambda^{(1)}\tau}\ee^{-\lambda^{(2)}\tau}\cdots
  \ee^{-\lambda^{(n)}\tau},\nonumber\\
  &= \ee^{-(\lambda^{(1)} + \lambda^{(2)} + \cdots
    + \lambda^{(n)})\,\tau}.
\end{align}
We may rewrite this as
\begin{equation}
  \theta \sim \ee^{-\Lambda_n \time},
  \label{eq:meanlambda}
\end{equation}
where~$\time=n\tau$, and
\begin{equation*}
  \Lambda_n \ldef \frac{1}{n}\sum_{i=1}^n\lambda^{(i)}
\end{equation*}
is the `running' mean value of the stretching rate at the~$n$th period.  As we
let~$n$ become large, how de we expect the concentration field to decay?  We
might expect that it would decay at the mean value~$\bar\lambda$ of the
stretching rates~$\lambda^{(i)}$.  This is not the case: the running
mean~\eqref{eq:meanlambda} does not \emph{converge} to the mean~$\bar\lambda$.
Rather, by the central limit theorem its expected value is~$\bar\lambda$, but
its fluctuations around that value are proportional to~$1/\sqrt{\time}$.
These fluctuations have an impact on the decay rate of~$\theta$.

\comment{Independent variables for CLT.}

The ensemble of variables~$\lambda^{(i)}$ is known as a \emph{realisation}.
Now let us imagine performing our blob experiment several times, and averaging
the resulting concentration fields: this is known as an \emph{ensemble
average} over realisations.  Ensemble-averaging smooths out fluctuations
present in each given realisation.  We may then replace the running
mean~$\Lambda_n$ by a sample-space variable~$\Lambda$, together with its
probability distribution~$P(\Lambda,\time)$.  The mean (expected value) of the
$\alpha$th power of the concentration field is then proportional to
\begin{equation}
  \Expect{\theta^\alpha} \sim \int_0^\infty \ee^{-\alpha\Lambda\time}\,
  \Prob(\Lambda,\time)\dint\Lambda\,.
  \label{eq:Expectthetaalpha}
\end{equation}
The overbar denotes the expected value.  The
factor~$\ee^{-\alpha\Lambda\time}$ gives the amplitude of~$\theta^\alpha$
given that the mean stretching rate at time~$\time$ is~$\Lambda$,
and~$\Prob(\Lambda,\time)$ measures the probability of that value of~$\Lambda$
occuring at time~$\time$.

The form of the probability distribution function (PDF)~$\Prob(\Lambda,\time)$
is given by the central limit theorem:
\begin{equation}
  \Prob(\Lambda,\time) \simeq
  \Gauss\bigl(\Lambda-\bar\Lambda;\sqrt{\fluc/\time}\bigr),
  \label{eq:ProbGaussian}
\end{equation}
that is, a Gaussian distribution~\eqref{eq:Gaussdef} with mean~$\bar\Lambda$
and standard deviation~$\sqrt{\fluc/\time}$.  The quantity~$\fluc$ is a
measure of the nonuniformity (or fluctuations) of stretching in the flow, both
spatially and temporally.  The decrease of the standard
deviation~$\sqrt{\fluc/\time}$ with time reflects the convergence of the
average stretching rate to the Lyapunov exponent~$\bar\Lambda$.  I will say
more about~$\fluc$ in Section~\ref{sec:decayexample}, when we look at a
practical example.

Actually, the central limit theorem only applies to values of~$\Lambda$ that
do not deviate too much from the mean.  The theorem understimates the
probability of rare events; a more general form of the PDF of~$\Lambda$ comes
from \emph{large deviation theory}~\cite{Ellis,ShwartzWeiss},
\begin{equation}
  \Prob(\Lambda,\time) \simeq
  \sqrt{\frac{\time\,\Cramer''(0)}{2\pi}}\,
  \ee^{-\time\Cramer(\Lambda-\bar\Lambda)}.
  \label{eq:ProbLargeDev}
\end{equation}
(A derivation of~\eqref{eq:ProbLargeDev} is given in
Appendix~\ref{sec:largedev}.)  The function~$\Cramer(x)$ is known as the
\emph{rate function}, the \emph{entropy function}, or the \emph{Cram\'er
function}, depending on the context (that is, which literature one is
reading).  It is a time-independent convex function with a minimum value
of~$0$ at~$0$: \hbox{$\Cramer(0) = \Cramer'(0) = 0$}.  If~$\Lambda$ is near
the mean, we have
\begin{equation}
  \Cramer(\Lambda-\bar\Lambda) \simeq
  \tfrac{1}{2}\,\Cramer''(0)(\Lambda-\bar\Lambda)^2,
  \label{eq:Gapprox}
\end{equation}
which recovers the Gaussian result~\eqref{eq:ProbGaussian}
with~\hbox{$\fluc=1/\Cramer''(0)$}.  Both~\eqref{eq:ProbGaussian}
and~\eqref{eq:ProbLargeDev} are only valid for large~$\time$ (which in our
case means~\hbox{$\time\gg\tau$}, or equivalently~\hbox{$n\gg1$}).

\comment{More on choosing which form: really depends on which part of the
  integral dominates.  If too far from the mean, then must use LDT.}

We can now evaluate the integral~\eqref{eq:Expectthetaalpha} with the
PDF~\eqref{eq:ProbLargeDev},
\begin{equation}
  \Expect{\theta^\alpha} \sim 
  \int_0^\infty \ee^{-\time H(\Lambda)}\,\dint\Lambda
  \sim \ee^{-\gamma_\alpha\time}\,,
  \label{eq:theint}
\end{equation}
where we have omitted the nonexponential prefactors, and defined
\begin{equation*}
  H(\Lambda) \ldef \alpha\Lambda + \Cramer(\Lambda-\bar\Lambda).
\end{equation*}
Since~$\time$ is large, the integral is dominated by the minimum value of
$H(\Lambda)$: this is the perfect setting for the well-known saddle-point
approximation.  The minimum occurs at~$\Lambda_\spt$
where~$H'(\Lambda_\spt)=\alpha + \Cramer'(\Lambda_\spt-\bar\Lambda)=0$, and is
unique because~$\Cramer$ is convex and has a unique minimum. \comment{See
27/05/2004 notes.}  The decay rate is then given by
\begin{equation}
  \gamma_\alpha = H(\Lambda_\spt),\quad
  \text{with\quad } H'(\Lambda_\spt)=0.
  \label{eq:gamma1}
\end{equation}

There's a caveat to this: for~$\alpha$ large enough the saddle
point~$\Lambda_\spt$ is negative.  This is not possible: the
stretching rates are defined to be nonnegative (the integral~\eqref{eq:theint}
involves only nonnegative~$\Lambda$).  Hence, the best we can do is to
choose~$\Lambda_\spt=0$---the integral~\eqref{eq:theint} is dominated
by realisations with no stretching.  Thus, in that
case~\hbox{$\gamma_\alpha=H(0)$}, or
\begin{equation}
  \gamma_\alpha = \Cramer(-\bar\Lambda).
  \label{eq:gamma2}
\end{equation}
We re-emphasise: for small enough~$\alpha$, the saddle point is positive and
the decay rate is given by~\eqref{eq:gamma1}.  Beyond that, we must choose
zero as the saddle point and the decay rate is given by~\eqref{eq:gamma2}.  To
find the critical value~$\alpha_{\mathrm{crit}}$ where we pass
from~\eqref{eq:gamma1} to~\eqref{eq:gamma2}, observe that this happens as the
saddle point nears zero.  Thus, we may solve our saddle point
equation~$H'(\Lambda_\spt)=0$ by Taylor expansion,
\begin{equation}
  H'(\Lambda_\spt) \simeq \alpha_{\mathrm{crit}}
  + \Cramer'(-\bar\Lambda)
  + \Lambda_\spt\,\Cramer''(-\bar\Lambda) = 0.
  \label{eq:spsmall}
\end{equation}
But the saddle point will not be small unless the first terms cancel
in~\eqref{eq:spsmall}, that is~\hbox{$\alpha_{\mathrm{crit}} =
-\Cramer'(-\bar\Lambda)$}.  We may thus recapitulate the result for the decay
rate,
\begin{equation}
  \gamma_\alpha = \begin{cases}
    \alpha\Lambda_\spt + \Cramer(\Lambda_\spt-\bar\Lambda),
    \qquad &\alpha < -\Cramer'(-\bar\Lambda);\\[6pt]
    \Cramer(-\bar\Lambda),\qquad &\alpha \ge -\Cramer'(-\bar\Lambda).
  \end{cases}
  \label{eq:gammanGauss}
\end{equation}
Clearly~$\gamma_\alpha$ is continous, and it can be easily shown
that~$d\gamma_\alpha/d\alpha$ is also continuous.

\begin{figure}
\psfrag{alpha}{\hspace{1em}\raisebox{-.3em}{$\alpha$}}
\psfrag{alphac}{\hspace{-.7em}\raisebox{.2em}{$\alpha_{\mathrm{crit}}$}}
\psfrag{gamma}{\hspace{.8em}\raisebox{.4em}{$\gamma_\alpha$}}
\centering
\includegraphics[width=.8\textwidth]{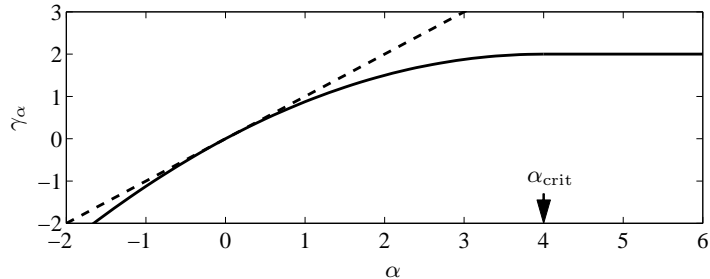}
\caption{Decay rate~\eqref{eq:gammaGauss} for the concentration of a blob
  in a Gaussian random stretching flow (solid curve).  The dashed line is for
  a fixed, nonrandom flow as in Section~\ref{sec:strain}.
  Here~$\bar\Lambda=1$,~$\fluc=1/4$,
  so~$\alpha_{\mathrm{crit}}=\bar\Lambda/\fluc=4$.}
\label{fig:gammablob}
\end{figure}

As an illustration, we use the Gaussian approximation~\eqref{eq:Gapprox} for
the Cram\'er function, with~$\fluc=1/\Cramer''(0)$.  The critical~$\alpha$
is~$\alpha_{\mathrm{crit}}=-\Cramer'(-\bar\Lambda)=\bar\Lambda/\fluc$.  The
saddle point is positive for~\hbox{$\alpha < \bar\Lambda/\fluc$}, so
from~\eqref{eq:gammanGauss} we get
\begin{equation}
  \gamma_\alpha = \begin{cases}\alpha\l(\bar\Lambda
    - \tfrac{1}{2}\,\alpha\fluc\r),\qquad &\alpha < \bar\Lambda/\fluc;\\[6pt]
    \bar\Lambda^2/2\fluc,\qquad &\alpha \ge \bar\Lambda/\fluc.
  \end{cases}
  \label{eq:gammaGauss}
\end{equation}
This is plotted in Figure~\ref{fig:gammablob}.
Notice that the solid curve (for a random flow) lies below the dashed line
(for a nonrandom flow).  This is a general result: if~$f(x)$ is a convex
function and~$x$ a random variable, \emph{Jensen's inequality} says that
\begin{equation}
  \Expect{f(x)} \ge f(\Expect{x}).
  \label{eq:Jensen}
\end{equation}
Now,~$\ee^{-\alpha\time\Lambda}$ is a convex funtion of~$\Lambda$, so we have
\begin{equation*}
  \Expect{\ee^{-\alpha\time\Lambda}} \ge \ee^{-\alpha\time\Expect{\Lambda}},
\end{equation*}
which means that the rate of decay satisfies
\begin{equation*}
  \gamma_\alpha \le \alpha\,\bar\Lambda,
\end{equation*}
which is exactly what is seen in Figure~\ref{fig:gammablob}.  Thus,
\emph{fluctuations in $\Lambda$ inevitably lead to a slower decay
rate~$\gamma_\alpha$}.

Stronger fluctuations also means that the decay rate~$\gamma_\alpha$ saturates
more quickly with~$\alpha$.  Clearly, in the absence of fluctuations we
recover the nonrandom result: $\bar\Lambda/\fluc$ is infinite and only
the~\hbox{$\alpha < \bar\Lambda/\fluc$} case is needed in
Eq.~\eqref{eq:gammaGauss}.  If there are lots of fluctuations,
$\bar\Lambda/\fluc$ is small, and there is a greater probability of obtianing a
realisation with no stretching.  For large enough fluctuations this
exponentially-decreasing probability dominates, and we obtain the second case
in~\eqref{eq:gammanGauss}.

\subsection{Many Blobs}
\label{sec:manyblobs}

\begin{figure}
\psfrag{(a)}{(a)}
\psfrag{(b)}{(b)}
\psfrag{(c)}{(c)}
\psfrag{(d)}{(d)}
\centering
\includegraphics[width=.8\textwidth]{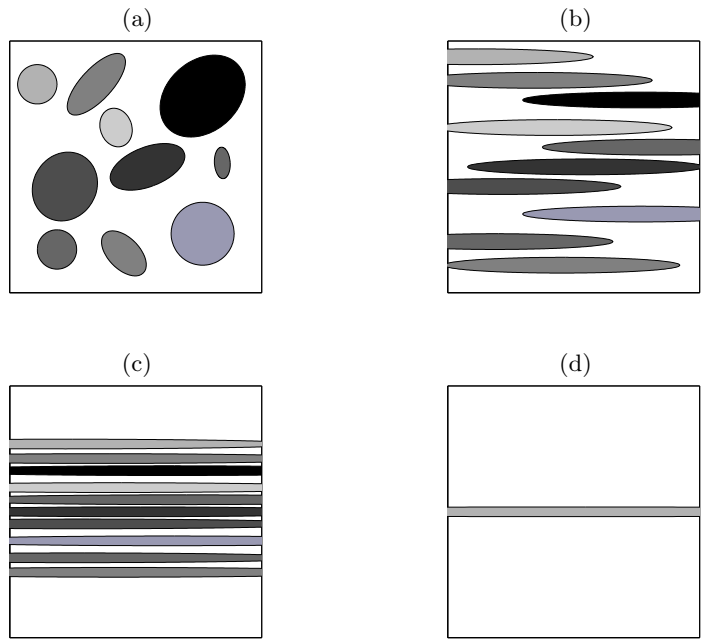}
\caption{(a) An initial distribution of blobs with random concentrations (b)
  are stretched by a constant strain (c) until they reach the diffusive limit
  in the contracting direction and begin to overlap.  (d) Finally, they
  combine into one very long blob with the average concentration of all the
  blobs.}
\label{fig:overlapblobs}
\end{figure}

In Section~\ref{sec:oneblob} we considered the evolution of the concentration
of a single blob of concentration in a random straining field.  Now we turn
our attention to a large number of blobs, homogeneously and isotropically
distributed, with random concentrations.  We assume that the mean
concentration over all the blobs is zero.  A simplified view of this initial
situation is depicted in Figure~\ref{fig:overlapblobs}(a), with shades of gray
indicating different concentrations.
If we now apply a uniform straining flow of the type (2a) (see
Section~\ref{sec:strain}), the blobs are all stretched horizontally
(the~$\xc_1$ direction) and contracted in the vertical ($\xc_2$) direction, as
shown in Figure~\ref{fig:overlapblobs}(b).  They are pressed together in
the~$\xc_2$ direction until diffusion becomes important
(Figure~\ref{fig:overlapblobs}(c)).  The effect of diffusion is to homogenise
the concentration field until it reaches a value which is the average of the
concentration of the individual blobs.  This is depicted by the long gray blob
in Figure~\ref{fig:overlapblobs}(d), which will itself keep contracting until
it reaches the diffusive length~$\fwid$.

Of course, 
here the initial concentration field~$\theta_0$ represents the concentration
of \emph{all} the homogeneously-distributed blobs together, so it does not
decay at infinity: we must thus use Eq.~\eqref{eq:thetastabilised2} rather
than~\eqref{eq:thetastabilised3}.  The summing (and hence averaging) over
blobs is manifest in Eq.~\eqref{eq:thetastabilised2}, which contains an
integral over the initial distribution~$\theta_0$ in the~$\xc_2$ direction,
windowed by a Gaussian.

In practice, this implies that the expected value of the concentration at a
point~$\xv$ on the gray filament is given by
\begin{equation}
  \l\langle\theta(\xv,\time)\r\rangle_{\mathrm{blobs}} \sim \ee^{-\Lambda\time}
  \sum_i^N \Expect{\theta_0^{(i)}} \longrightarrow 0\,,
  \label{eq:overlapblobs}
\end{equation}
where~$\theta_0^{(i)}$ is the initial concentration of the~$i$th blob,
and~$\l\langle\cdot\r\rangle_{\mathrm{blobs}}$ denotes the expected value of
the sum over the overlapping blobs at point~$\xv$ (not the same as spatial
integration~$\l\langle\cdot\r\rangle$).  We assume that~\hbox{$N\gg 1$} blobs
have overlapped.  Equation~\eqref{eq:overlapblobs} gives the concentration at
a point, summed over~$N$ overlapping blobs.  Of course,
Eq.~\eqref{eq:overlapblobs} converges to zero for large~$N$, because the blobs
average out.  Not so for the fluctuations at that point: by the central limit
theorem, we have
\begin{equation}
  \l\langle\theta^2(\xv,\time)\r\rangle_{\mathrm{blobs}}
  \sim \ee^{-2\Lambda\time} \sum_i^N \Expect{{\theta_0^{(i)}}^2}
  = N \ee^{-2\Lambda\time}\,\,\Expect{\theta_0^2}\,,
  \label{eq:overlapblobs2}
\end{equation}
since the initial blobs have identical distributions.  The blob-summed
fluctuation amplitude~$\l\langle\theta^2\r\rangle_{\mathrm{blobs}}$ is thus
proportional to the number~$N$ of overlapping blobs.  But the number of
overlapping blobs is proportional to~$\ee^{\Lambda\time}$: as time increases
more and more blobs converge to a given~$\xv$ in the contracting direction and
overlap diffusively (this can be seen in Eq.~\eqref{eq:thetastabilised2}: the
width of the windowing region grows as~$\ee^{\lambda\time}$).  Assuming the
variance of~$\theta_0^{(i)}$ is finite, we conclude
from~\eqref{eq:overlapblobs2} that
\begin{equation}
  \l\langle\theta^2(\xv,\time)\r\rangle_{\mathrm{blobs}}^{1/2}
  \sim \ee^{-\Lambda\time/2}\,.
  \label{eq:theta2manyblobs}
\end{equation}
Compare this to~\eqref{eq:meanlambda} for the single-blob case: the overlap
between blobs has led to an extra square root.  Thus, the ensemble
averages~$\Expect{\l\langle\theta^2(\xv,\time)
\r\rangle^{\alpha}}_{\!\!\!\mathrm{blobs}}$ for the overlapping blobs are
computed exactly as in Section~\ref{sec:oneblob}, resulting
in~\eqref{eq:gammanGauss}.  Because of the assumption of homogeneity, the
point-average is the same as the average over the whole domain (see
Section~\ref{sec:practical} for more on this), and we have\footnote{In going
from~\eqref{eq:theta2manyblobs} to~\eqref{eq:mommanyblobs}, we've implicitly
assumed that the initial concentration field has Gaussian statistics, because
we've used the fact that the higher even moments are proportional to powers of
the second moment.}
\begin{equation}
  \l\langle\theta^2\r\rangle^\alpha
  = \Expect{\l\langle\theta^2(\xv,\time)
    \r\rangle^{\alpha}}_{\!\!\!\mathrm{blobs}}
  \sim \ee^{-\gamma_\alpha\time}\,,
  \label{eq:mommanyblobs}
\end{equation}
with~$\gamma_\alpha$ given by~\eqref{eq:gammanGauss}.
(In~\eqref{eq:mommanyblobs} the angle brackets denote spatial averaging, not
spatial integration, because the total variance is infinite in this case.)

\subsection{Three Dimensions}
\label{sec:3Dblobs}

In three dimensions, we will only treat Case~(3a) (a purely straining flow) of
Section~\ref{sec:3D}.  For~$\lambda_2<0$, where the asymptotic concentration
is given by~\eqref{eq:ropes} (ropes), the situation is basically identical to
the 2D case of Sections~\ref{sec:oneblob}--\ref{sec:manyblobs}: the statistics
of the stretching direction~$\lambda_1$ determine~$\gamma_\alpha$
from~\eqref{eq:gammanGauss}.  The contracting directions~$\xc_2$ and~$\xc_3$
are stabilised by diffusion.

For~$\lambda_2\ge0$, the asymptotic concentration is given
by~\eqref{eq:pancakes} (pancakes).  We have two fluctuating quantities to
worry about ($\lambda_1$ and~$\lambda_2$).  But since the decay rate
in~\eqref{eq:pancakes} only depends on~$\lambda_3$, we can instead focus on
its fluctuations only.  For a single blob, the
average~\eqref{eq:Expectthetaalpha} is then replaced by
\begin{equation}
  \Expect{\theta^\alpha} \sim \int_0^\infty
  \ee^{-\alpha\lvert\Lambda_3\rvert\time}\,
  \Prob_3(\lvert\Lambda_3\rvert,\time)
  \dint\lvert\Lambda_3\rvert \sim \ee^{-\gamma_\alpha\time}\,,
  \label{eq:Expectthetaalpha3D}
\end{equation}
where of course~$\Lambda_3$ is the average of~$\lambda_3$.  This PDF achieves
a distribution of the large-deviation form~\eqref{eq:ProbLargeDev}.  The
analysis thus follows exactly as in Sections~\ref{sec:oneblob}
and~\ref{sec:manyblobs}, except the Cram\'er function
for~$\lvert\Lambda_3\rvert$ must be used.\footnote{There are a few exceptional
cases to consider~\cite{Balkovsky1999}.}

\section{Practical Considerations}
\label{sec:practical}

One may rightly wonder if the blobs in a random uniform straining flow
depicted in Section~\ref{sec:rsm} bear any resemblance to reality.  The
single-blob scenario doesn't, but the many-blobs scenario has a fighting
chance, as we will try to justify here.  There are two important
considerations: where does the ensemble-averaging come from, and what are the
stretching rates given by?

The decay rate~\eqref{eq:gammanGauss} depends crucially on ensemble-averaging:
with that averaging the decay rate fluctuates wildly for a given realisation.
At the end of Section~\ref{sec:manyblobs} we assumed that homogeneity allowed
us to generalise from the average at a point to the average over the whole
domain.  But the average over the whole domain can actually do a lot more for
us: it can provide the ensemble of blobs that we need for averaging!  Thus, we
can forget about speaking of realisations as if we were running many parallel
experiments, and instead speak of the moments of the concentration field as
given by an average over randomly-distributed blobs.  The decay rate will then
be naturally smoothed-out over blobs experiencing different stretching
histories.  The saturation of the decay rate with~$\alpha$ in
Eq.~\eqref{eq:gammanGauss} is due to~$\Expect{\theta^{2\alpha}}$ being
dominated by the fraction of blobs that have experienced no stretching.

What about the stretching rates~$\lambda$?  Luckily, it is not them but their
time-average~$\Lambda$ that matters.  If we imagine following a blob as it
moves through the flow, we can see that this time-averaged stretching rate is
nothing but the finite-time Lyapunov exponent associated with this blob and
its particular initial condition.  A given blob will be constantly reoriented
as it moves along in the flow, so its finite-time Lyapunov exponent is not
just the average of the stretching rates (in fact, it must be strictly less
than this average).  But in a chaotic system we are guaranteed that, on
average, these reorientations do not lead to a vanishing (infinite-time)
Lyapunov exponent.  This is guaranteed by the celebrated Oseledec
multiplicative theorem for random matrices~\cite{Oseledec1968}.%
\footnote{The reorientations also tend to decrease the correlation
time~$\tau$~\cite{Balkovsky1999}.}
We may thus use for~$\Prob(\Lambda,\time)$ the distribution of finite-time
Lyapunov exponents, which is well-known to have the large-deviation
form~\eqref{eq:ProbLargeDev}~\cite{Ott}.

The result of these considerations is the \emph{local theory} of passive
scalar decay.  It is called local because of the reliance of such a local
concept as the finite-time Lyapunov exponents, which come from a linearisation
near fluid element trajectories.  In Section~\ref{sec:decayexample} we discuss
a specific example.  We postpone a discussion of the validity of the local
theory until Section~\ref{sec:limitations}, but for now we point out that it
is known to be exact at least in some simple
model flows~\cite{Fouxon1998,Balkovsky1999}.

The derivation presented in this Section was based on the work of Balkovsky
and Fouxon~\cite{Balkovsky1999}, who used a slightly more rigorous approach.
Son~\cite{Son1999} also obtained the decay rate~\eqref{eq:gammanGauss} using
path-integral methods.  Earlier, Antonsen~\etal~\cite{Antonsen1996} derived
the decay rate for the second moment~$ \l\langle\theta^2\r\rangle$ in terms
of the Cram\'er function, using a different (and not quite equivalent)
approach, though they did not allow for the second case
in~\eqref{eq:gammanGauss}.

\subsection{An Example: Flow in a Microchannel}
\label{sec:decayexample}

\begin{figure}
\psfrag{x}[c][c]{\Large{$x$}}
\psfrag{y}[c][c]{\Large{$y$}}
\psfrag{z}[c][c]{\Large{$z$}}
\psfrag{0}[c][c]{\Large{$0$}}
\psfrag{h}[c][c]{\Large{$h$}}
\psfrag{-l/2}[c][c]{\hspace{-1ex}\large{$-\frac{\ell}{2}$}}
\psfrag{l/2}[c][c]{\large{$\frac{\ell}{2}$}}
\centering
\includegraphics[width=.7\textwidth]{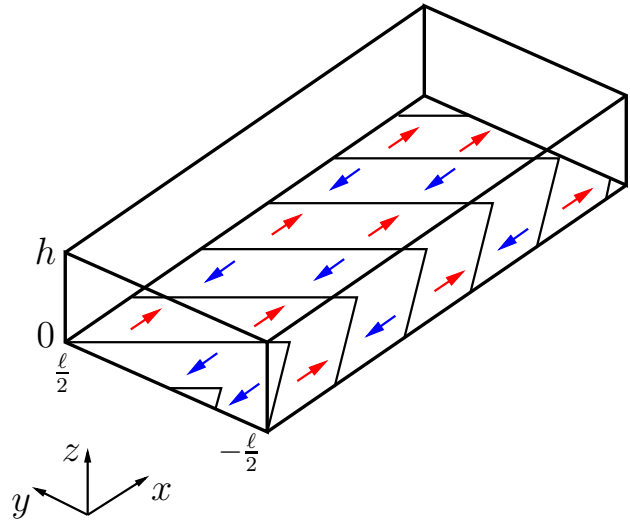}
\caption{Microchannel with a periodic patterned electro-osmotic potential at
the bottom.  The arrows indicate the direction of fluid motion at the bottom.
The width of the channel is about~$100\, \mu$m and its height $10$--$50\,
\mu$m, and the period of the pattern is~$L$.  A typical mean fluid velocity
is~$10^2$--$10^3\, \mu\mathrm{m}/\mathrm{s}$.}
\label{fig:microchannel}
\end{figure}

We illustrate how to compute the decay rates~$\gamma_\alpha$ with a practical
problem.  Specifically, we will use a three-dimensional model of a
microchannel.  The system is shown in Figure~\ref{fig:microchannel}.
It consists of a narrow channel, roughly~$100\,\mu\mathrm{m}$ wide and
slightly shallower.  These types of channel are widely used in microfluidics
applications (``lab-on-a-chip''), and often one wants to achieve good mixing
in the lateral cross-section of the channel.  This is difficult, since the
Reynolds number of the flow varies between~$0.1$ and~$100$---far from
turbulent.  Clever techniques have to be used to induce chaotic motion of the
fluid particle trajectories in order to enhance mixing.
Stroock~\etal~\cite{Stroock2002} used patterned grooves at the bottom of the
channel to induce vortical motions, and found that the mixing efficiency was
dramatically increased.  Here we use a variation on this where the bottom is
pattern with an electro-osmotic coating, which induces fluid motion near the
wall~\cite{HongIMECE2003}.  The effect of the electro-osmotic coating is
well-approximated by a moving wall boundary condition.  The pattern is chosen
in a so-called herringbone pattern to maximise the mixing efficiency (though
not in a staggered herringbone, which is even better but is more difficult to
model).  Rather than solving the full equations numerically, we adopt here an
analytical model based on Stokes flow in a shallow
layer~\cite{EwartThiffeaultPreprint}.  The longitudinal ($x$) direction is
taken to be periodic.  The flow is steady, but because it is three-dimensional
it can still exhibit chaos.

\begin{figure}
\psfrag{y}[c][c]{\raisebox{-1.5ex}{$y$}}
\psfrag{z}[c][c]{\raisebox{0.9ex}{$z$}}
\psfrag{Section a3b5, a=0, U0=5, Lx0.000}[c][c]{Section at $x=0$}
\psfrag{Section a3b5, a=0, U0=5, Lx0.500}[c][c]{Section at midpoint $x=L/2$}
\centering
\includegraphics[width=.7\textwidth]{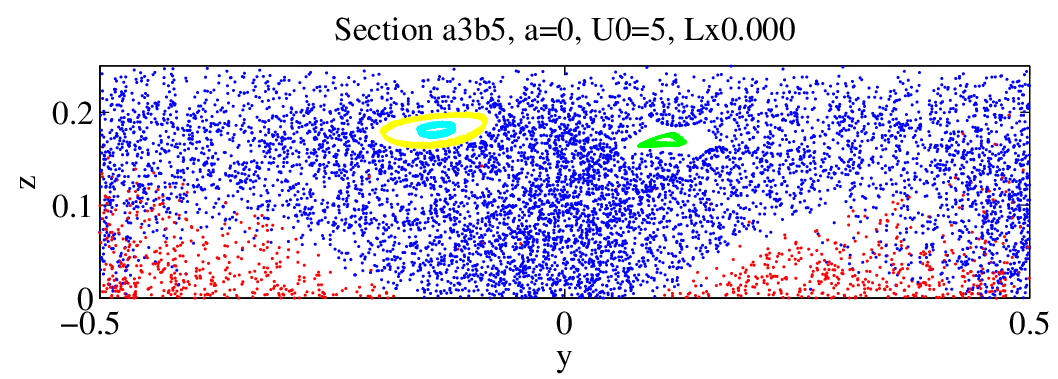}

\vspace*{.5em}

\centering
\includegraphics[width=.7\textwidth]{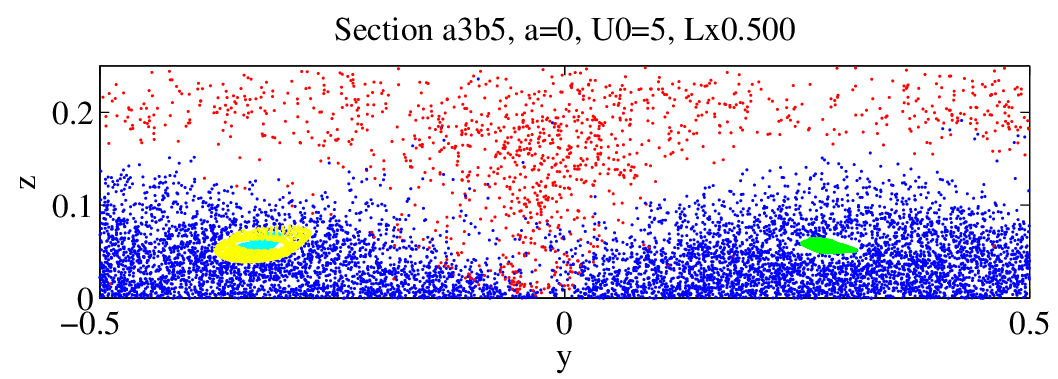}

\caption{Poincar\'e sections for the microchannel.  The red and blue dots
  represent the same trajectory periodically puncturing two vertical
  planes many times over (blue if in the same direction as the flow, red
  otherwise).  The green and yellow dots show two trajectories in regular,
  nonmixing regions.}
\label{sec:Psections}
\end{figure}

Figure~\ref{sec:Psections} shows two Poincar\'e sections for the flow.  These
are taken at two constant~$x$ planes, one at~$x=0$ and the other at the
mipoint of the~$x$-periodic pattern.  The two colours represent two
trajectories that have periodically punctured those planes many times over.
It is clear from the Figures that the flow contains large chaotic regions, as
well as smaller regular regions (known as islands).  We focus here on the
chaotic regions.

\begin{figure}
\psfrag{PDF of FTLEs}{\hspace{-.5em}\raisebox{.5em}{\tiny PDF of FTLEs}}
\psfrag{FTLEs}{\hspace{1em}\raisebox{-.5em}{\hspace{-1em}\tiny FTLEs}}
\begin{center}
\begin{minipage}{.45\textwidth}
\psfrag{U0=5 with 20 modes}
       {\raisebox{.25em}{\hspace{.5em}$\time=19\,\mathrm{s}$}}
\centering
\includegraphics[width=.8\textwidth]
	{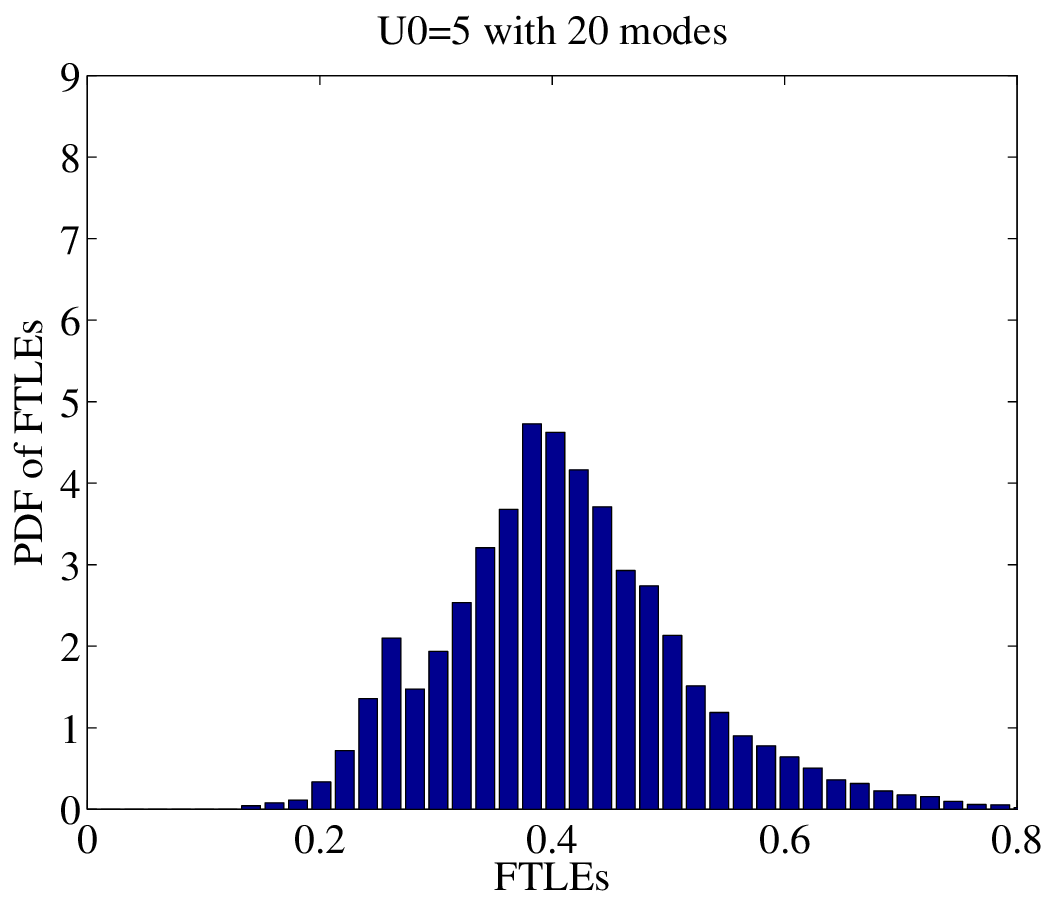}
\end{minipage}
\begin{minipage}{.45\textwidth}
\psfrag{U0=5 with 20 modes}
       {\raisebox{.25em}{\hspace{.5em}$\time=38\,\mathrm{s}$}}
\hspace{1em}
\centering
\includegraphics[width=.8\textwidth]
	{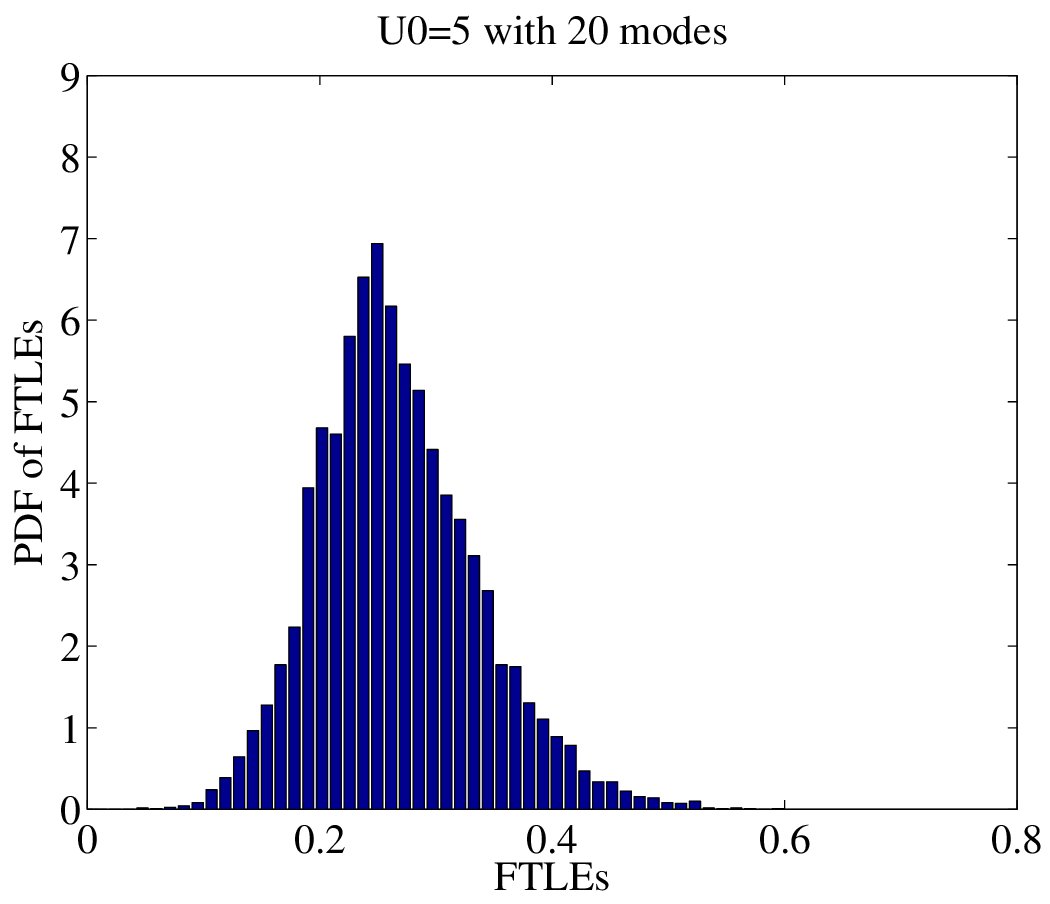}
\end{minipage}

\vspace{.25in}

\begin{minipage}{.45\textwidth}
\psfrag{U0=5 with 20 modes}
       {\raisebox{.25em}{\hspace{.5em}$\time=57\,\mathrm{s}$}}
\centering
\includegraphics[width=.8\textwidth]
	{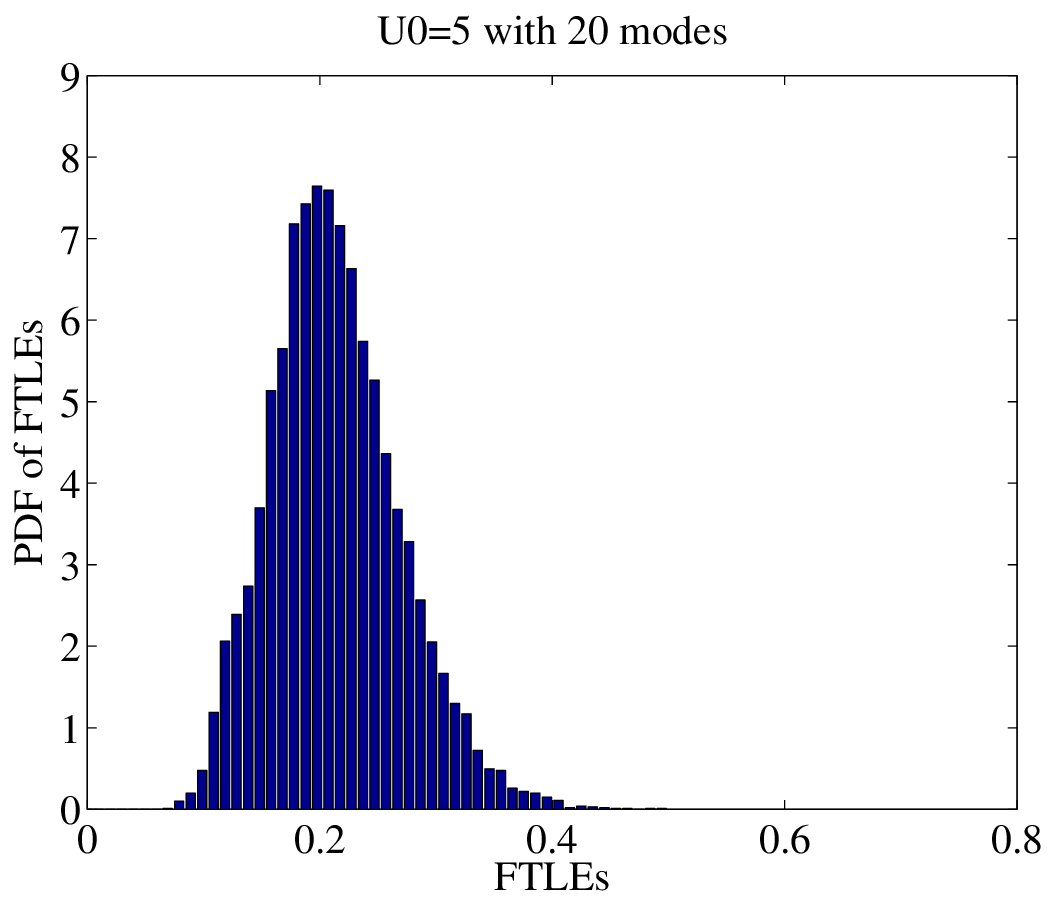}
\end{minipage}
\begin{minipage}{.45\textwidth}
\hspace{1em}
\psfrag{U0=5 with 20 modes}
       {\raisebox{.25em}{\hspace{.5em}$\time=76\,\mathrm{s}$}}
\centering
\includegraphics[width=.8\textwidth]
	{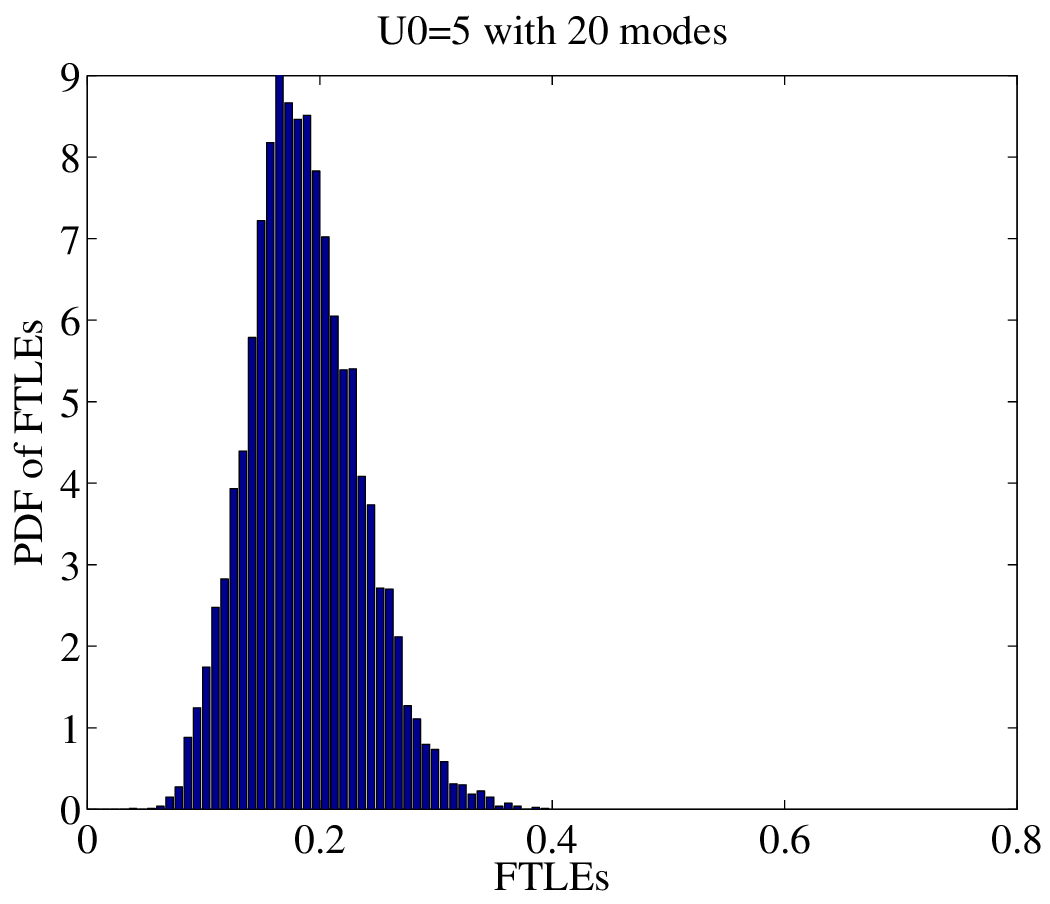}
\end{minipage}
\end{center}
\caption{Evolution of the distribution of finite-time Lyapunov exponents for
  the microchannel. The average crossing time for particles in the
  channel is~$L/U$.}
\label{fig:Lyapdist}
\end{figure}

Now that we have established (or at least strongly suspect) the existence of
chaotic regions, we can compute the distribution of finite-time Lyapunov
exponents.  There are many ways of doing this: because we are not interested
in exetremely long times, the most direct route may be used.  We have an
analytical form for the velocity field, so the velocity gradient matrix is
easily computed.  This allows us to linearise about trajectories in the
standard manner~\cite{Eckmann1985,Ott}.  Each trajectory will thus have a
finite-time Lyapunov exponent associated with it, which shows the tendency of
infinitesimally close trajectories to diverge exponentially.  This is then
repeated over many different trajectories within the same chaotic region, and
a histogram is made of the finite-time Lyapunov exponents.  This histogram
changes with time, as shown in Figure~\ref{fig:Lyapdist}.
For these relatively early times, it changes dramatically and does not show
a self-similar form.

\begin{figure}
\begin{center}
\subfigure[]{
  \psfrag{Mean}{\hspace{-1em}\raisebox{.5em}{Mean}}
  \psfrag{Time}{\raisebox{-.6em}{$\time$}}
  \psfrag{Mean of 1st Lyapunov Exponent vs Time}{}
  \centering
  \includegraphics[width=.44\textwidth]
		  {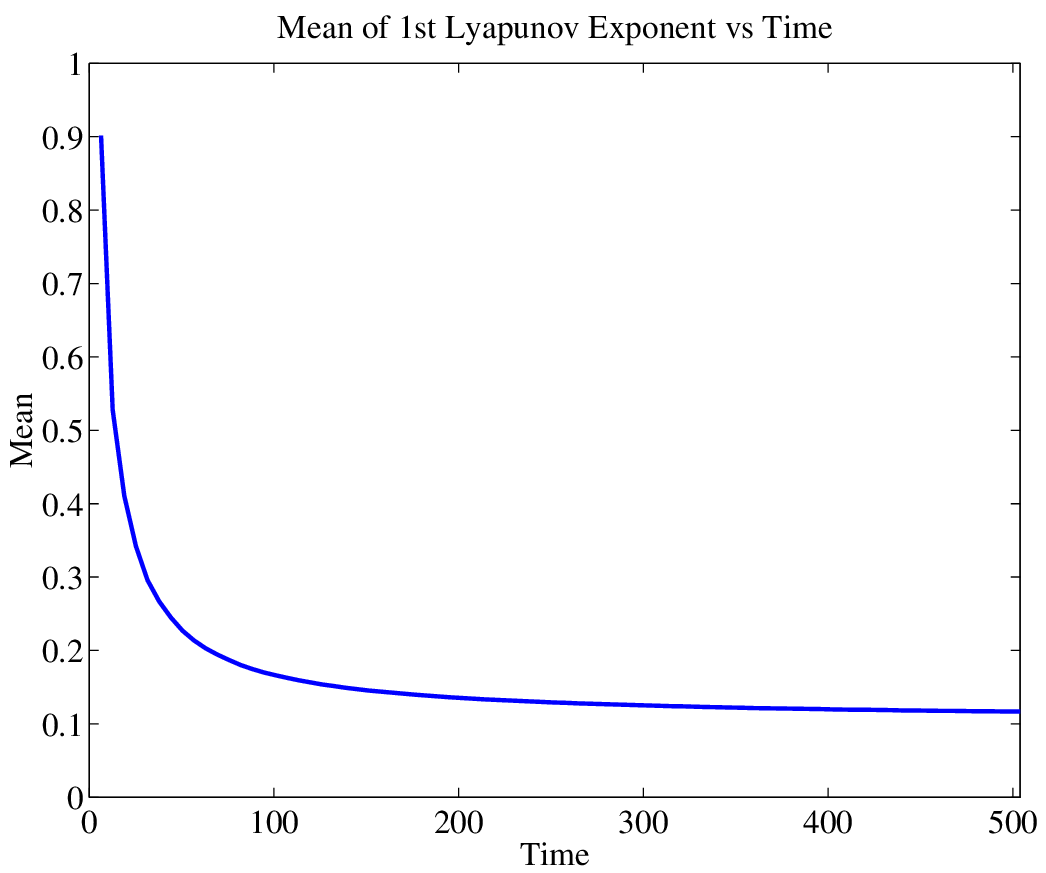}
}\goodgap%
\subfigure[]{
  \psfrag{Standard Deviation}
	 {\hspace{-2.5em}\raisebox{.5em}{Standard Deviation}}
  \psfrag{1/sqrt(Time)}{\raisebox{-.6em}{$1/\sqrt{\time}$}}
  \psfrag{Standard Deviation of Lyapunov Exponent vs 1/sqrt(Time)}{}
  \centering
  \includegraphics[width=.44\textwidth]
		  {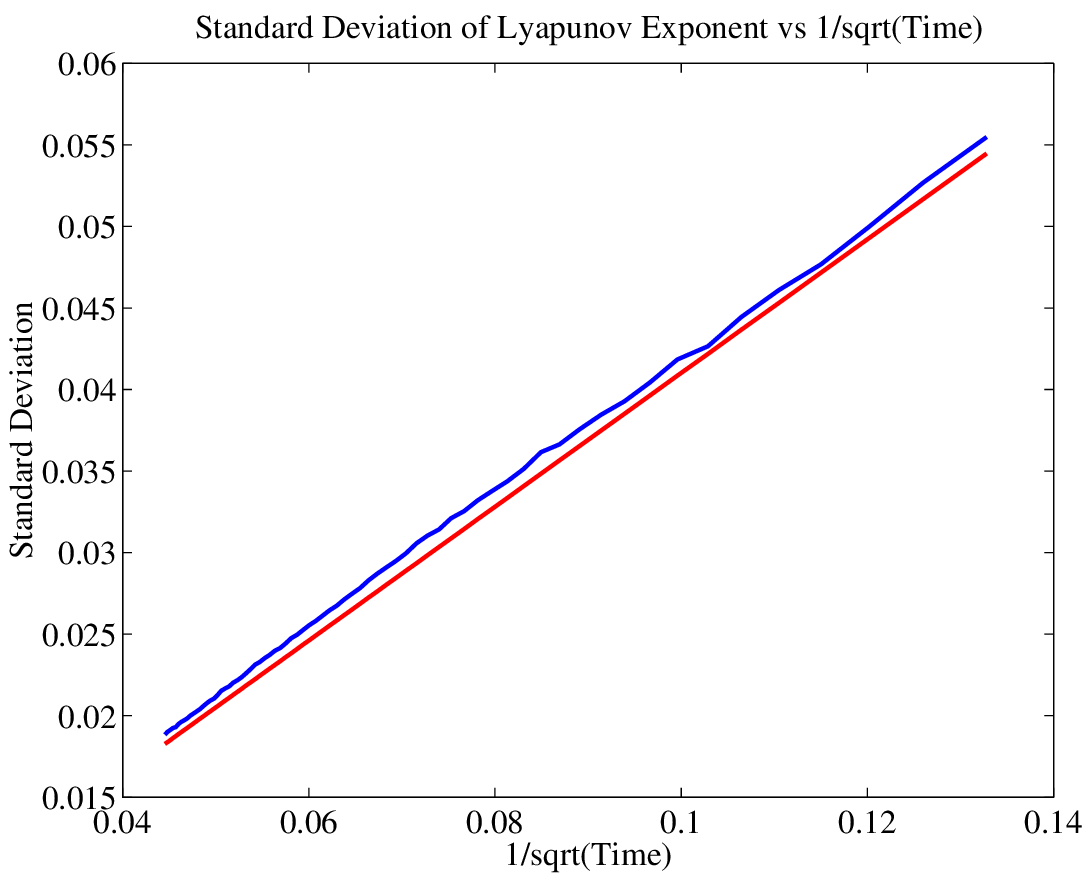}
}
\end{center}
\caption{(a) Evolution of the mean~$\bar\Lambda$ of the distribution of
  Lyapunov exponents.  The mean converges to~$\bar\Lambda \simeq 0.116\,\,
  {\mathrm{s}}^{-1}$.  (b) Standard deviation of the distribution of Lyapunov
  exponents versurs~$1/\sqrt{\time}$.  The straight line represents
  $\sqrt{{\fluc}/{\time}}$, with $\fluc \simeq 0.168\,\, {\mathrm{s}}^{-1}$.}
\label{fig:Lyapdistmeanstd}
\end{figure}

\begin{figure}
\psfrag{Shifted Lyapunov Exponents with 20 Fourier Modes}{}
\psfrag{\(Lyapunov Exponent - Mean\) * sqrt\(Time\)}
       {\hspace{5.75em}\raisebox{-.5em}{$\sqrt{\time}\,(\Lambda-\bar\Lambda)$}}
\psfrag{Normalised PDF}{\hspace{-.5em}\raisebox{.25em}{Normalised PDF}}
\centering
\includegraphics[width=.7\textwidth]
		{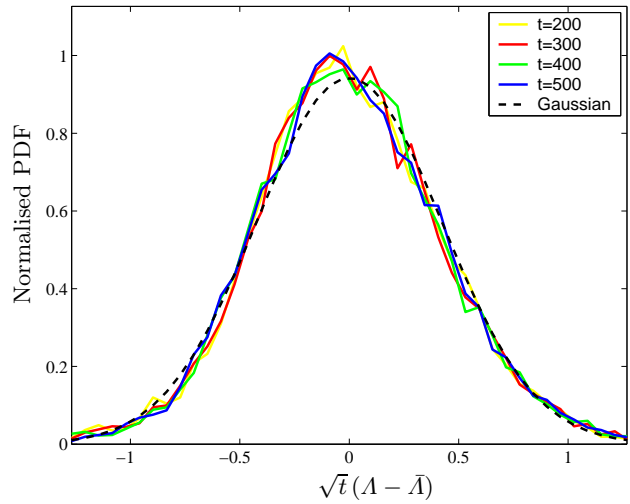}
\caption{Rescaled distribution of finite-time Lyapunov exponents at different
times.  The dashed line is the Gaussian form~\eqref{eq:ProbGaussian}, with
parameters as in the caption for Figure~\ref{fig:Lyapdistmeanstd}.}
\label{fig:Lyapdistresc}
\end{figure}

The evolution of the mean and standard deviation of the distribution is shown
in Figure~\ref{fig:Lyapdistmeanstd}.  The mean is converging to a
constant~\hbox{$\bar\Lambda \simeq 0.116$}, and the standard deviation is
decreasing as~$\sqrt{\fluc/\time}$, with~\hbox{$\fluc \simeq 0.168$}.
\setlength{\subfigcapskip}{4pt}%
\setlength{\subfigtopskip}{0pt}%
\setlength{\subfigbottomskip}{0pt}%
(Both~$\bar\Lambda$ and~$\fluc$ are fitted values, and arise from the
complicated nonlinear nature of fluid particle trajectories---they cannot be
predicted or calculated from first principles except in the simplest cases.)
These facts taken together are strongly indicative that the distribution is
converging to a Gaussian of the form~\eqref{eq:ProbGaussian}.  This is easily
confirmed by plotting the PDFs at different times and rescaling the horizontal
axis by~$\sqrt{\time}$, as shown in Figure~\ref{fig:Lyapdistresc}.
Note that this case exhibits a particularly nice Gaussian form, which is not
necessarily the norm for all chaotic flows.

Using the values for~$\bar\Lambda$ and~$\fluc$ we just obtained, we can
calculate the decay rates~$\gamma_\alpha$ with the Gaussian approximation.
The ratio~$\bar\Lambda/\fluc$ is~$0.69$, so the change in character in
\eqref{eq:gammaGauss} occurs at~\hbox{$\alpha \ge 0.69$}.  Since
from~\eqref{eq:mommanyblobs} the decay of~$\l\langle\theta^2\r\rangle^\alpha$
is given by~$\gamma_\alpha$, this means that moments of order~\hbox{$2\alpha
\ge 1.38$} will decay at the same rate.  This includes the
variance~$\l\langle\theta^2\r\rangle$, so we have from~\eqref{eq:mommanyblobs}
\begin{equation*}
  \l\langle\theta^2\r\rangle
  \sim \ee^{-\gamma_1\time},\quad\text{with}\quad
  \gamma_1 = \bar\Lambda^2/2\fluc
  \simeq 0.040\,\, {\mathrm{s}}^{-1}.
\end{equation*}
The mixing time is thus~\hbox{$\gamma_1^{-1} \simeq 25$} seconds.  This is
about a factor of four improvement over the purely diffusive time for, say,
DNA molecules ($\Diff \simeq 10^{-10}\,\,\mathrm{m^2\,s}^{-1}$).  This is not
spectacular, but can be greatly increased by staggering the herringbone
pattern.  The mixing time assuming the decay proceeds at the rate of the mean
Lyapunov exponent~$\bar\Lambda$ is roughly~$9$ seconds, so that the
fluctuations multiply this by a factor of three!

Of course, we do not know if this is actually a good estimate for the mixing
time, since we haven't directly solved the \adeq\ numerically: this is
prohibitive in a three-dimensional domain for such a small diffusivity.  This
is one of the advantages of the local theory: it is usually less expensive to
compute the distribution of finite-time Lyapunov exponents than it is to solve
the \adeq\ equation directly.  We will say more on the validity of the local
theory in Section~\ref{sec:limitations}.

\subsection{Limitations of the Local Theory}
\label{sec:limitations}

So is this local theory of mixing correct?  Well, certainly not always, even
in the Batchelor regime.  There are many assumptions underlying the model,
some of them difficult to verify. (Do blobs really undergo a series of
stretching events as described here?  Do correlations between these events
matter?)  My feeling is that sometimes it will, but most of the time it won't.
More experiments and numerical simulations are needed to get to the bottom of
this.  For a detailed discussion of possible problems with the local theory,
see Fereday and Haynes~\cite{Fereday2004}.  They make a good case that the
theory must break down for long times: the blobs discussed here meet the
boundaries of the fluid domain and must begin to fold.  The folding forces
them to interact with themselves in a correlated fashion.  We enter the regime
of the \emph{strange eigenmode}~\cite{Pierrehumbert1994}, which has received a
lot of attention lately~\cite{Rothstein1999,Fereday2002,%
Sukhatme2002,Wonhas2002,Pikovsky2003,Thiffeault2003d,Fereday2004,%
Liu2004,Thiffeault2004b,Schekochihin2004}.  Maybe we'll hear more about that
in ten years\dots.%
\footnote{Recently, Tsang~\etal~\cite{Tsang2005} have tested the local
  theories to an astonishing precision for the Zeldovich sine flow, whilst
  Haynes and Vanneste~\cite{Haynes2005} have convincingly demonstrated that
  the local theory holds when the system is dominated by the remnants of the
  continuous spectrum of the advection operator, whereas global aspects must
  be considered when the slowest-decaying mode is regular.}

\section*{Acknowledgments}

I thank Martin Ewart, Bill Young, Andy Thompson, and Emmanuelle Gouillart for
their valuable comments, as well as the organisers of the Aosta school,
Antonello Provenzale and Jeff Weiss.

\appendix

\section{The Advection--Diffusion Equation in a Comoving Frame}
\label{sec:adeqcomov}

\mathnotation{\rc}{r}
\mathnotation{\rv}{\bm{\rc}}

We start from the \adeq~\eqref{eq:adeq} and derive its
form~\eqref{eq:adeqlinv} for a linearised velocity field.  We want to
transform from the fixed spatial coordinates~$\xv$ to coordinates~$\rv$
measured from a reference fluid trajectory~$\xv_0(\time)$.  The
coordinates~$\rv$ are not quite material (Lagrangian) coordinates, since we
follow the trajectory of only one fluid element.

We thus let
\begin{equation}
  \xv = \xv_0(\time) + \rv,
  \qquad
  \frac{d\xv_0(\time)}{d\time} = \velv(\xv_0(\time),\time),
  \label{eq:reftraj}
\end{equation}
and write the concentration field as
\begin{equation*}
  \theta(\xv,\time) = \tilde\theta(\rv,\time).
\end{equation*}
The time derivative of~$\theta$ can be written
\begin{equation}
  {\l.\frac{\pd}{\pd\time}\r\rvert}_{\xv}\theta(\xv,\time)
  = {\l.\frac{\pd}{\pd\time}\r\rvert}_{\rv}\tilde\theta(\rv,\time)
  + \grad_{\rv}\tilde\theta
  \cdot{\l.\frac{\pd\rv}{\pd\time}\r\lvert}_{\xv},
  \label{eq:changeth}
\end{equation}
where~$\pd/\pd\time\rvert_{\xv}$ denotes a derivative with~$\xv$ held
constant.  Now from~\eqref{eq:reftraj}
\begin{equation*}
  {\l.\frac{\pd\rv}{\pd\time}\r\lvert}_{\xv}
  = -\frac{d\xv_0}{d\time} = -\velv(\xv_0(\time),\time).
\end{equation*}
Spatial derivatives are unchanged by~\eqref{eq:reftraj}: $\grad_{\xv}\theta =
\grad_{\rv}\tilde\theta$.  Hence, inserting~\eqref{eq:changeth}
into~\eqref{eq:adeq}, we find
\begin{equation}
  {\l.\frac{\pd}{\pd\time}\r\rvert}_{\rv}\tilde\theta
  + \{\velv(\xv_0(\time) + \rv,\time)-\velv(\xv_0(\time),\time)\}
  \cdot\grad_{\rv}\tilde\theta
  = \Diff\,\lapl_{\rv}\tilde\theta\,.
  \label{eq:adeqcomov}
\end{equation}
We Taylor expand the velocity field in~\eqref{eq:adeqcomov} to get
\begin{equation}
  {\l.\frac{\pd}{\pd\time}\r\rvert}_{\rv}\tilde\theta
  + \rv\cdot\sigma(\time)\cdot\grad_{\rv}\tilde\theta
  = \Diff\,\lapl_{\rv}\tilde\theta\,,
  \qquad \sigma(\time) \ldef \grad\velv(\xv_0(\time),\time),
  \label{eq:adeqcomov2}
\end{equation}
where we neglected terms of order~$\lvert\rv\rvert^2$.  This is only valid if
the velocity field changes little over the region we consider (\ie, if it is
smooth enough), which is true for large Schmidt number.
Equation~\eqref{eq:adeqcomov2} is the same as~\eqref{eq:adeqlinv}, and tells
us how to find~$\sigma(\time)$.

\section{Volume Preservation}
\label{sec:detT1}

It is useful to know how to prove that a divergence-free vector field will
lead to volume-preservation of a small blob of integrated trajectories.
Mathematically, we want to go from~\eqref{eq:Tfevol} to~\eqref{eq:detT1}.  We
start from the definition of the determinant of a~$\sdim$-dimensional matrix,
\begin{equation}
  \det\Tf = \sum_{i_1\cdots i_\sdim}\sum_{j_1\cdots j_\sdim}
  \frac{1}{\sdim!}\,\epsilon_{i_1\cdots i_\sdim}\,
  \epsilon_{j_1\cdots j_\sdim}\,\Tf_{i_1 j_1}\cdots\Tf_{i_\sdim j_\sdim}
  \label{eq:detT}
\end{equation}
where we have dropped the `$\time$' subscript from~$\Tf$ for this
section, and~$\epsilon$ is the fully-antisymmetric Levi-Civita symbol.
Taking a time derivative of~\eqref{eq:detT},
\begin{equation}
  \pd_\time\det\Tf = \sum_{i_1\cdots i_\sdim}\sum_{j_1\cdots j_\sdim}
  \frac{1}{(\sdim-1)!}\,\epsilon_{i_1\cdots i_\sdim}\,
  \epsilon_{j_1\cdots j_\sdim}\,
  \pd_\time\Tf_{i_1 j_1}\cdots\Tf_{i_\sdim j_\sdim}
\end{equation}
since the~$\sdim$ terms obtained after using the product rule for derivatives
are the same.  We use the equation of motion~\eqref{eq:Tfevol} for~$\Tf$,
\begin{equation*}
  \pd_\time\det\Tf
  = -\sum_{i_1\cdots i_\sdim}\sum_{j_1\cdots j_\sdim}\sum_\ell
  \frac{1}{(\sdim-1)!}\,\epsilon_{i_1\cdots i_\sdim}\,
  \epsilon_{j_1\cdots j_\sdim}\,
  \sigma_{i_1 \ell}\,\Tf_{\ell j_1}\,
  \Tf_{i_2 j_2}\cdots\Tf_{i_\sdim j_\sdim}\,.
\end{equation*}
We rearrange the sums on the right,
\begin{equation*}
  \pd_\time\det\Tf
  = -\sum_{i_1,j_1,\ell}
  \sigma_{i_1 \ell}\,\Tf_{\ell j_1}\,
  \l(\sum_{i_2\cdots i_\sdim}\sum_{j_2\cdots j_\sdim}
  \frac{1}{(\sdim-1)!}\,\epsilon_{i_1\cdots i_\sdim}\,
  \epsilon_{j_1\cdots j_\sdim}\,
  \Tf_{i_2 j_2}\cdots\Tf_{i_\sdim j_\sdim}\r)
\end{equation*}
and recognise that, up to a factor of~$\det\Tf$, the term in the parentheses
is the cofactor representation of the inverse of~$\Tf$:
\begin{equation*}
  \pd_\time\det\Tf
  = -\sum_{i_1,j_1,\ell}
  \sigma_{i_1 \ell}\,\Tf_{\ell j_1}\,(\Tf^{-1})_{j_1i_1}\,\det\Tf
  = -\sum_{i_1,\ell}
  \sigma_{i_1 \ell}\,\delta_{\ell i_1}\,\det\Tf
\end{equation*}
so that finally
\begin{equation*}
  \pd_\time\det\Tf
  = -(\Tr\sigma)\,\det\Tf\,.
\end{equation*}
We conclude that if~$\Tr\sigma=0$ then the determinant of~$\Tf$ is constant.
Since its initial condition is unity (from Eq.~\eqref{eq:Tfevol}), then it
must remain so for all time.

\section{Large Deviation Theory}
\label{sec:largedev}

In this Appendix we will justify the large-deviation form of the PDF,
Eq.~\eqref{eq:ProbLargeDev}, assuming little prior knowledge of probability
theory.

First, we define the \emph{characteristic function} $\ee^{-s(k)}$ of a random
variable~$x$ by
\begin{equation*}
  \ee^{-s(k)} = \int p(x)\,\ee^{-\imi\,kx}\,\dint x\,,
\end{equation*}
that is, the characteristic function is simply the Fourier transform of the PDF
of~$x$.  We have~$s(0)=0$,~$\bar x = -\imi\,s'(0)$, and~$\overline{x^2} -
{\bar x}^2 = s''(0)$.  Now define the random variable~$X$ to be the mean of
$n$ variables,
\begin{equation*}
  X_n = \frac{1}{n}\,\sum_{i=1}^n x_i
\end{equation*}
where the $x_i$ are independent and identically distributed with
PDF $p(x_i)=p(x)$.  How do we find the PDF $P(X_n)$ of $X_n$, in the
limit where~$n$ is large?  First observe that (from here on we drop the
subscript on $X_n$)
\begin{equation*}
  P(X)
  = \int p(x_1)\cdots p(x_n)\,\,\delta\l(\frac{x_1+\cdots+x_n}{n}-X\r)
  \!\dint x_1\cdots\dint x_n
\end{equation*}
since the joint PDF $p(x_1,\ldots,x_n) = p(x_1)\cdots p(x_n)$ by
independence of the $x_i$.  The characteristic function $\ee^{-S(k)}$ for $P(X)$
is then
\begin{align*}
  \ee^{-S(k)} &= \int P(X)\,\ee^{-\imi k X}\dint X\nonumber\\
  &= \int p(x_1)\cdots p(x_n)\,
  \delta\!\l(\frac{x_1+\cdots+x_n}{n}-X\r)
  \!\ee^{-\imi k X}\dint x_1\cdots\dint x_n\dint X
\end{align*}
We do the $X$ integral, and then observe that we get a product of~$n$
identical $x_i$ integrals, each of which is equal to~$\ee^{-s(k/n)}$:
\begin{equation*}
  \ee^{-S(k)} = \int p(x_1)\cdots p(x_n)\,\,
  \ee^{-\imi (k/n) (x_1 + \cdots + x_n)}\dint x_1\cdots\dint x_n
  = \ee^{-n s(k/n)}\,.
  \label{eq:PX}
\end{equation*}
Thus the characteristic function for $X$ is the $n$th power of the characteristic
function for $x$.  We can invert the Fourier transform to find the PDF $P(X)$:
\begin{equation}
  P(X) = \frac{1}{2\pi}\int\ee^{-S(k)}\,\ee^{\imi\,kX}\dint k
  = \frac{n}{2\pi}\int\ee^{-n(s(K) - \imi\,KX)}\dint K
  \label{eq:PX2}
\end{equation}
where $K=k/n$.  We let
\begin{equation*}
  H(K,X) = s(K) - \imi\,KX\,;
\end{equation*}
For large values of $n$ the integral in~\eqref{eq:PX2} is dominated by the
stationary points~$K_\spt(X)$ of $H(K,X)$ (saddle-point approximation):
\begin{equation}
  K_\spt(X) \quad\text{such that}\quad
  \frac{\pd H}{\pd K}(K_\spt,X)=s'(K_\spt) - \imi\,X = 0.
  \label{eq:KstpX}
\end{equation}
\comment{Only one?  Convexity?}  (I will often leave out the $X$ dependence of
$K_\spt(X)$ to shorten the expressions.)  In that case we can approximate the
integrand in~\eqref{eq:PX2} using
\begin{equation*}
  H(K,X) = H(K_\spt,X) + \tfrac{1}{2}\,s''(K_\spt)(K-K_\spt)^2
  + \Order{(K-K_\spt)^3}
\end{equation*}
which allows us to do the integral explicitly:
\begin{equation}
  P(X) = \sqrt{\frac{n}{2\pi\, s''(K_\spt(X))}}\,\,\ee^{-nH(K_\spt(X),X)}
  \label{eq:PXlargedev}
\end{equation}
where~$K_\spt(X)$ is given by~\eqref{eq:KstpX}.  As a final step, let us
calculate the mean of $X$ using this PDF:
\begin{equation}
  \overline{X} = \int X P(X) \dint X
  = \int X \sqrt{\frac{n}{2\pi\, s''(K_\spt(X))}}\,\,\ee^{-nH(K_\spt(X),X)}
  \dint X\,.
  \label{eq:Xbar}
\end{equation}
Again, for large~$n$ we can use the saddle-point method to evaluate this
integral.  The important observation is that the saddle-point $X_\sptt$
of~$H_\spt(K(X),X)$ satisfies
\begin{equation*}
  \frac{dH}{dX}(K_\spt(X_\sptt),X_\sptt) = \frac{\pd
  H}{\pd K}(K_\spt(X_\sptt),X_\sptt)
  \,\frac{dK_\spt}{dX}(X_\sptt) -\imi\,K_\spt(X_\sptt) = 0.
\end{equation*}
The~\hbox{$\pd H/\pd K$} term vanishes because it is evaluated at~$K_\spt$;
hence,~$K_\spt(X_\sptt)=0$, which implies~$H(K_\spt(X_\sptt),X_\sptt)=0$.
Inserting this into the integral~\eqref{eq:Xbar}, we find~$\overline X =
X_\sptt$: \emph{the mean of $X$ and the minimum of~$H$ coincide}.  This means
that it makes sense to define
\begin{equation}
  \Cramer(X - \overline X) \ldef H(K_\spt(X),X),\quad
  \text{with } \Cramer(0) = 0 \text{ and } \Cramer'(0) = 0,
  \label{eq:Cramerdef}
\end{equation}
which is the sought-after Cram\'er function.  Note also
that~$\Cramer''(X-\overline X) = 1/s''(K_\spt(X))$,%
\comment{
\[
S(X - \overline X) = s(K_\spt(X)) - \imi\,K_\spt(X)X
\]
\[
S'(X - \overline X) = s'(K_\spt(X))K_\spt'(X)
- \imi\,K_\spt'(X)X - \imi\,K_\spt(X)
\]
\begin{align*}
S''(X - \overline X) &= s''(K_\spt(X))(K_\spt'(X))^2
+ s'(K_\spt(X))K_\spt''(X) - \imi\,K_\spt''(X)X - 2\imi\,K_\spt'(X)\\
&= s''(K_\spt(X))(K_\spt'(X))^2 - 2\imi\,K_\spt'(X)\\
&= -\imi\,K_\spt'(X)\\ &= 1/s''(K_\spt(X))
\end{align*}
}
and that for large~$n$ the non-exponential coefficient
in~\eqref{eq:PXlargedev} can thus be approximated by evaluating it at the
saddle-point~$K_\spt(\overline X)=0$, with $s''(0)=1/\Cramer''(0)$.  The final
form of our large-deviation result is thus
\begin{equation}
  P(X) = \sqrt{\frac{n\,\Cramer''(0)}{2\pi}}\,\,\ee^{-n\Cramer(X-\overline X)}
  \label{eq:PXlargedev2}
\end{equation}
which is the same as Eq.~\eqref{eq:ProbLargeDev}.

As a simple example (treated in every textbook, see for
example~\cite{ShwartzWeiss}), consider a random variable~$x$ with PDF
\begin{equation}
  p(x) = (1-\varepsilon)\,\delta(x-x_+) + \varepsilon\,\delta(x-x_-),
  \label{eq:binom}
\end{equation}
where~$x_+ > x_-$ are constants---this is a \emph{binomial distribution} (or
Bernoulli distribution in this case).  If we take the mean~$X$ of~$n$ such
variables, what is the PDF of~$X$ for large~$n$?  First, we compute the
characteristic function for~$x$,
\begin{align}
  \ee^{-s(k)} &= \int \l\{(1-\varepsilon)\,\delta(x-x_+) +
  \varepsilon\,\delta(x-x_-)\r\}\ee^{-\imi\,kx}\dint x\nonumber\\
  &= (1-\varepsilon)\,\ee^{-\imi\,kx_+} + \varepsilon\,\ee^{-\imi\,kx_-}\,.
\end{align}
We take the logarithm to obtain~$s(k)$ and find~$K_\spt(X)$ by solving
the saddle-point equation~\eqref{eq:KstpX},
\begin{equation*}
  \frac{\pd H}{\pd K} = s'(K_\spt) - \imi\,X = 0
  \quad\Longleftrightarrow\quad
  K_\spt(X) = \frac{1}{\imi\Delta}\,
  \log\l(\frac{1-\varepsilon}{\varepsilon}\,\frac{x_+ - X}{X - x_-}\r),
\end{equation*}
where~\hbox{$\Delta \ldef x_+ - x_-$} and we restrict~$x_- \le X \le x_+$.
Inserting this into $H(K_\spt(X),X)$, we find from~\eqref{eq:Cramerdef}
\begin{equation*}
  \Cramer(X-\overline{X}) = -\frac{X - x_-}{\Delta}\,
  \log\l(\frac{1-\varepsilon}{\varepsilon}\,\frac{x_+ - X}{X - x_-}\r)
  + \log\l(\frac{x_+ - X}{\varepsilon\Delta}\r).
\end{equation*}
It is easy to verify that, since~\hbox{$\overline{X} = (1-\varepsilon)x_+ +
\varepsilon\,x_-$}, we have~$\Cramer(0)=\Cramer'(0)=0$ and~$\overline{X^2} -
{\overline X}^2 = 1/\Cramer''(0) = \varepsilon(1-\varepsilon)\,\Delta^2$.

The binomial distribution~\eqref{eq:binom} is a useful model of stretching of
an infinitesimal line segment by a uniform incompressible straining flow in
two dimensions, assuming the straining axes of the flow change direction
randomly at regular intervals~$\tau$.  If we set~$x_\pm=\pm\lambda\tau =
\pm\fluc$, where~$\lambda$ is the strain rate, then~$X$ is the averaged
logarithm of the length~$\ell$ of the segment, \ie\ $\ell = \ee^{nX}$.  Thus,
the $m$th power of the length of the segment will on average grow as
\begin{equation}
  \overline{\ell^m}
  = \overline{\ee^{mnX}} = \ee^{-S(\imi\,mn)} = \ee^{-n s(\imi\,m)}
  = \l\{(1-\varepsilon)\,\ee^{\fluc m} + \varepsilon\,\ee^{-\fluc m}\r\}^n\,.
  \label{eq:lm}
\end{equation}
We know that~$\overline{\ell^{-2}}$ must be constant in a 2D incompressible
flow~\cite{Falkovich2001}, so that the term in braces in~\eqref{eq:lm} must be
unity.  We use this to solve for~$\varepsilon$,
\begin{equation*}
  \varepsilon = (1 + \ee^{2\fluc})^{-1},
\end{equation*}
which then allows us to use~\eqref{eq:lm} to write the growth rate~$\chi_m$ of
line segments as
\begin{equation*}
  \chi_m = \frac{1}{\tau\,n}\log\overline{\ell^m}
  = \frac{1}{\tau}\log\l(\frac{\cosh(m+1)\fluc}{\cosh\fluc}\r)
\end{equation*}
The Lyapunov exponent, which is given by~$d\chi_m/dm$ at~$m=0$, has a value
of~$\lambda\tanh\fluc$ for this flow: it is less than for a uniform straining
flow because of the time taken for the segment to realign with the new
straining axis when its direction changes.


\end{document}